\documentclass[iop]{emulateapj}

\shorttitle{Resolving the Galaxies within a Giant \lya\ Nebula}

\shortauthors{Prescott et al.}

\begin{document}
\newcommand{\lya}{\hbox{{\rm Ly}\kern 0.1em$\alpha$}}
\newcommand{\heii}{\hbox{{\rm He}\kern 0.1em{\sc ii}}}
\newcommand{\civ}{\hbox{{\rm C}\kern 0.1em{\sc iv}}}
\newcommand{\ergss}{erg~s$^{-1}$}
\newcommand{\ergscmarc}{erg~s$^{-1}$~cm$^{-2}$~arcsec$^{-2}$} 
\newcommand{\ergscm}{erg~s$^{-1}$~cm$^{-2}$} 
\newcommand{\bw}{$B_{W}$}
\newcommand{\ib}{$IA445$}
\newcommand{\zblob}{z$\approx$2.7} 

\newcommand{\lyafactor}{42.9} 
\newcommand{\heiifactor}{22.9} 
\newcommand{\apsize}{0.4} 
\newcommand{\diffapsize}{1.6} 
\newcommand{\apcorrV}{1.15} 
\newcommand{\apcorr}{[1.15, 1.52, 1.76]} 
\newcommand{\maxheii}{1.0} 
\newcommand{\maxheiikpc}{8.0} 
\newcommand{\maxheiikpcr}{8} 
\newcommand{\minheii}{0.58} 
\newcommand{\minheiikpc}{4.6} 
\newcommand{\minheiikpcr}{5} 
\newcommand{\numgalagn}{17}
\newcommand{\numgal}{16}
\newcommand{\nummone}{9}
\newcommand{\nummtwo}{8}
\newcommand{\agnid}{36}
\newcommand{\lbgidA}{26}
\newcommand{\lbgidB}{59}
\newcommand{\interidA}{46}
\newcommand{\interidB}{60}
\newcommand{\specbotidA}{1}
\newcommand{\specbotidB}{57}
\newcommand{\spectopidA}{21}
\newcommand{\spectopidB}{58}
\newcommand{\lbgdist}{2.1} 
\newcommand{\lyauvoffset}{0.8} 
\newcommand{\numtrials}{100} 
\newcommand{\overdenA}{4}  
\newcommand{\overdensigA}{1}  
\newcommand{\overdenB}{12}  
\newcommand{\overdensigB}{3}  
\newcommand{\overdenC}{15}  
\newcommand{\overdensigC}{5}  
\newcommand{\degree}{\ensuremath{^\circ}}

\title{\sc{Resolving the Galaxies within a Giant \lya\ Nebula: Witnessing the Formation of a Galaxy Group?}}

\author{Moire K. M. Prescott\altaffilmark{1,2}, Arjun Dey\altaffilmark{3}, 
Mark Brodwin\altaffilmark{4,5}, Frederic H. Chaffee\altaffilmark{6}, Vandana Desai\altaffilmark{7}, Peter Eisenhardt\altaffilmark{8}, 
Emeric Le Floc'h\altaffilmark{9}, Buell T. Jannuzi\altaffilmark{3}, Nobunari Kashikawa\altaffilmark{10}, 
Yuichi Matsuda\altaffilmark{11}, B. T. Soifer\altaffilmark{7,12}} 

\altaffiltext{1}{TABASGO Postdoctoral Fellow; Department of Physics, University of California, Santa Barbara, CA 93106, USA; mkpresco@physics.ucsb.edu}
\altaffiltext{2}{Steward Observatory, University of Arizona, 933 N. Cherry Avenue, Tucson, AZ 85721, USA}
\altaffiltext{3}{National Optical Astronomy Observatory, 950 North Cherry Avenue, Tucson, AZ 85719, USA} 
\altaffiltext{4}{Harvard-Smithsonian Center for Astrophysics, 60 Garden Street, Cambridge, MA 02138, USA}  
\altaffiltext{5}{Department of Physics, University of Missouri - Kansas City, 5110 Rockhill Road, Kansas City, MO 64110, USA}
\altaffiltext{6}{Department of Physics and Astronomy, University of North Carolina, Chapel Hill, NC, 27599, USA; currently at Gemini Observatory} 
\altaffiltext{7}{Spitzer Science Center, California Institute of Technology, MS 220-6, Pasadena, CA 91125, USA} 
\altaffiltext{8}{Jet Propulsion Laboratory, California Institute of Technology, MC 169-327, 4800 Oak Grove Drive, Pasadena, CA 91109, USA} 
\altaffiltext{9}{Laboratoire AIM, CEA/DSM, CNRS, University of Paris, 91191, France} 
\altaffiltext{10}{Optical and Infrared Astronomy Division, National Astronomical Observatory of Japan, Mitaka, Tokyo, 181-8588, Japan} 
\altaffiltext{11}{Department of Physics, Science Site, Durham University, South Road, Durham, DH1 3LE, UK} 
\altaffiltext{12}{Caltech Optical Observatories, California Institute of Technology, Pasadena, CA 91125, USA} 

\begin{abstract}

Detailed analysis of the substructure of Ly$\alpha$ nebulae can put important constraints on the physical 
mechanisms at work and the properties of galaxies forming within them.  
Using high resolution HST imaging of a \lya\ nebula at $z\approx2.656$, we have taken 
a census of the compact galaxies in the vicinity, used optical/near-infrared colors to select system members, 
and put constraints on the morphology of the spatially-extended emission.  
The system is characterized by (a) a population of compact, 
low luminosity ($\sim0.1~L^{*}$) sources --- \numgalagn\ primarily young, small ($R_{e}\approx1-2$~kpc), disky galaxies 
including an obscured AGN --- that are all substantially offset ($\gtrsim$20~kpc) 
from the line-emitting nebula; (b) the lack of a central galaxy at or near the peak of the \lya\ emission; and (c) 
several nearly coincident, spatially extended emission components --- \lya, \heii, and UV continuum --- that are 
extremely smooth.  These morphological findings are difficult to reconcile with theoretical models that 
invoke outflows, cold flows, or resonant scattering, suggesting that while all of these physical phenomena may be occurring, 
they are not sufficient to explain the powering and large extent of \lya\ nebulae.  
In addition, although the compact galaxies within the system are irrelevant as power sources, the region 
is significantly overdense relative to the field galaxy population (by at least a factor of \overdenA).  
These observations provide the first estimate of the luminosity function of galaxies within an individual \lya\ nebula 
system, and suggest that large \lya\ nebulae may be the seeds of galaxy groups or low-mass clusters.  

\end{abstract}

\keywords{galaxies: evolution --- galaxies: formation --- galaxies: high-redshift}

\section{Introduction}
\label{sec:intro}

Giant ($\sim$100~kpc), radio-quiet \lya\ nebulae (or \lya\ ``blobs") that have been discovered 
in the distant universe by virtue of their extremely luminous \lya\ emission ($\sim10^{44}$\ergss) 
are thought to be regions of ongoing massive galaxy formation.  
When studied in detail, these systems show complex morphologies, obscured active galactic 
nuclei (AGN) and/or associated star-forming galaxies 
\citep[e.g.,][]{francis96,ivi98,stei00,pal04,chap04,mat04,mat07,basu04,dey05,gea07,smi08,pres09,ouc09,yang11a}.  
There is strong evidence that the largest Ly$\alpha$ nebulae are rare 
\citep{sai06,yang09,yang10,presphd,mat11} and typically reside in the most overdense regions 
of the Universe \citep[e.g.,][]{pal04,mat04,mat05,mat09,sai06,pres08,yang09}.  

Unlike \lya\ halos observed around quasars and radio galaxies (e.g., \citealt{mcc93}, and 
references therein; \citealt{wei05,miley06,barrio08,smith09}), the dominant power source 
responsible for these radio-quiet \lya\ nebulae has been difficult to determine.  
Studies have investigated whether \lya\ nebulae could be powered by galactic superwind 
outflows \citep[e.g.,][]{tani00,tani01,mori04}, 
photoionization by obscured AGN or star formation \citep[e.g.,][]{chap04,basu04,gea07,gea09,pres09}, 
or gravitational cooling within cold filaments \citep[``cold flows"; e.g.,][]{nil06,smi07,dijkloeb09,goerdt10,faucher10}, 
but have come to a range of conclusions.  It has also been suggested that the large extent of 
\lya\ nebulae could be due to resonant scattering of \lya\ photons from a central source, with 
recent work providing observational evidence for this effect around Lyman-break galaxies \citep{stei11}.  
Thus, despite considerable study, the mechanisms responsible for the copious \lya\ emission 
and the large extent of \lya\ nebulae have remained controversial.  

The question of the substructure of \lya\ nebulae has received much less attention, but it is a topic that can provide 
much-needed complementary clues to the origin of \lya\ nebulae as well as to what they ultimately evolve into.  
As the potential physical mechanisms responsible for powering the \lya\ emission each have morphological implications, 
studying the morphology of the spatially-extended emission on kiloparsec and sub-kiloparsec scales can provide 
insight into the processes at work and the underlying powering mechanism in \lya\ nebulae.  
In addition, taking a complete census of the diffuse and compact sources 
within \lya\ nebulae and studying the relative positions and properties (luminosities, colors, morphologies, sizes) 
of the galaxies that reside within them is valuable for establishing the evolutionary state of these dynamic systems.
Doing all of this, however, requires determining the membership of these crowded regions either with spectroscopy 
(typically feasible only for the brightest knots) or deep, high resolution imaging that can resolve and put 
constraints on the spectral energy distribution (SED) of faint individual sources.  As yet, very few radio-quiet 
\lya\ nebulae have been imaged with HST, and those that have typically lack the critical high resolution constraints 
above the Balmer break. 

In this paper we study the sub-kiloparsec structure of a giant \lya\ nebulae at $z\approx2.656$ using 
high-resolution imaging from HST.  This \lya\ nebula was discovered thanks to its extreme {\it Spitzer}/MIPS 24\micron\ 
emission and its extended morphology in broad-band \bw\ imaging 
\citep[LABd05;][hereinafter Paper~I]{dey05}.  Roughly 20\arcsec\ ($\sim$160 kpc) in size with a 
\lya\ luminosity of $\approx$1.7$\times$10$^{44}$ erg s$^{-1}$, LABd05 rivals other known \lya\ nebulae 
in energetics and complexity.  
The data presented in Paper~I revealed at least three important components (and potential sources of ionization) 
in the system: (1) the strong 24\micron\ source, likely dominated by an 
obscured AGN, (2) a Lyman break galaxy to the northeast of the nebula, 
and (3) a source that does not have a counterpart in the ground-based imaging but 
that was identified near the center of the \lya\ emission 
due to the presence of narrow, spatially unresolved \heii$\lambda$1640 and \civ$\lambda$1548,1550 emission lines 
in the ground-based spectrum. 
Follow-up narrow-band imaging of the surrounding environment revealed that LABd05 resides within a 
very large filamentary structure at least 50 comoving Mpc in size \citep{pres08}, and imaging 
polarization observations have demonstrated that the \lya\ emission is not strongly polarized 
\citep[$P<9\%$, $3\sigma$;][]{pres11}, 
hinting that scattering may not be significant in this source.  

While revealing, our previous studies were limited by ground-based resolution and depth, and 
our resulting knowledge of the system was incomplete.  
First, at ground-based resolution, it was unclear whether the system hosted other compact galaxies 
and whether the \lya\ itself contained a compact central source or was clumpy on small scales.  
Second, it was unclear whether the 24\micron\ source and the Lyman break galaxy 
were important power sources for the \lya\ nebula.  
The geometry of the system, with both the 24\micron\ source and the Lyman break galaxy 
offset from the centroid of the \lya\ by 2\farcs5 ($\gtrsim$20~kpc in projection), argued 
against this possibility, and the observed SED of both sources suggested that, barring 
inhomogeneous obscuration, they were unlikely to power more than $\sim$20\% of 
the \lya\ emission.  
Finally, the source of the unresolved \heii\ and \civ\ emission, 
which appeared to be centered within the \lya\ emission, was uncertain.  
While \heii\ and \civ\ emission often indicate shock excitation or a hard ionization source, 
no central galaxy (that could be driving shock-heating via a superwind) was visible 
in the ground-based imaging and furthermore the measured line ratios were inconsistent with shocks.

In the present work, we use high resolution HST/ACS and NICMOS imaging 
to take a census of the compact sources within the system, measure their luminosities, morphologies, and 
locations relative to the line-emitting gas, investigate the question of the location and morphology 
of the \heii-emitting region, and determine the morphology of the \lya\ nebula itself.  
In Section~\ref{sec:obsredux} we describe our observations and reductions, and 
in Section~\ref{sec:components} we present our results on the different components of the \lya\ 
nebula system.  Section~\ref{sec:discussion} summarizes what we have learned about the small-scale morphology 
of LABd05 and explores the implications of these findings for our understanding of what causes the 
\lya\ nebula phenomenon.  We conclude in Section~\ref{sec:conclusions}.  
In this paper, we assume the standard $\Lambda$CDM cosmology ($\Omega_{M}$=0.3, $\Omega_{\Lambda}$=0.7, $h$=0.7);
the angular scale at $z=2.656$ is 7.96 kpc/\arcsec.  All magnitudes are in the AB system \citep{oke74}.

\section{Observations \& Reductions}
\label{sec:obsredux}

Paper~I examined the properties of LABd05 using Keck/LRIS spectroscopy, optical imaging from 
the NOAO Deep Wide-Field Survey \citep[NDWFS;][]{jan99}, and Spitzer/IRAC and MIPS imaging \citep{eisen04,houck05}.  
The large-scale environment of LABd05 was studied using Subaru/SuprimeCam intermediate-band \ib\ 
\lya\ imaging \citep{pres08}, and constraints on the \lya\ polarization of the system were obtained 
using imaging polarimetry with the Bok Telescope and the SPOL CCD Spectropolarimeter \citep{pres11}.  
In this paper, we add high resolution HST/ACS and NICMOS imaging to study the small-scale morphology 
and local environment of LABd05.  
Table~\ref{tab:hstobs} lists the instruments, filters, and total exposure times for the 
HST imaging.  Figure~\ref{fig:imall} shows selected postage stamps from the multi-wavelength dataset 
used in this work.

\subsection{HST ACS Data}
\label{sec:acs}

We obtained HST Advanced Camera for Surveys (ACS) imaging of LABd05 on UT 2006 January 13, 14, and 24 
using the $F606W$ ($V_{606}$) filter and two 2\% ramp filters, $FR462N$ (centered on \lya\ at \zblob) and 
$FR601N$ (centered on \heii$\lambda$1640 at \zblob).\footnote{HST Cycle 14; GO 10591}  
Basic image calibrations (overscan, bias, and dark subtraction, flat-fielding) were provided by the standard 
HST ACS pipeline with On-The-Fly-Reprocessing (OTFR) and the task {\it calacs}.  
We removed a residual offset in the bias level of the individual amplifiers 
on each of the ACS detectors (roughly a 2\% effect relative to the background) by 
estimating the sky background in each amplifier separately using a sigma-clipped mean 
and subtracting it from the calibrated, flat-fielded individual exposures (the ``\_FLT" files).  
Using MultiDrizzle's default settings and no additional sky subtraction, we performed the distortion correction, 
cosmic-ray rejection, and image combination, yielding a final scale of 0.05\arcsec/pixel and 
a field-of-view of 207\arcsec$\times$205\arcsec.  
The point-spread-function (PSF) size for the ACS imaging is FWHM=0.07\arcsec, as measured using the
TinyTim PSF emulator.\footnote{TinyTim: http://www.stsci.edu/hst/observatory/focus/TinyTim.}  
The 5$\sigma$ point source limiting magnitudes for the ACS imaging are 28.3, 25.7, 
and 25.7 mag (\apsize\arcsec\ diameter aperture) for the F606W, FR462N, and F601N filters, respectively.  

The narrow-band \lya\ and \heii\ imaging contain both line and continuum emission.  
To generate HST line-only \lya\ and \heii\ images, we scaled the $V_{606}$ image by factors of 
\lyafactor\ and \heiifactor, respectively 
(estimated empirically based on sources common to both images) and subtracted the result from 
the original narrow-band images.  The resulting HST line-only \lya\ and \heii\ images are 
discussed in Sections~\ref{sec:lya} and \ref{sec:heii}.  
To quantify the degree to which the \heii\ and \civ\ emission lines in turn contaminate the $V_{606}$-band, 
we measured the total $V_{606}$ flux within the same spectroscopic aperture ($4.5\arcsec\times1.5$\arcsec) 
used in Paper~I to measure the \heii\ and \civ\ fluxes, after convolving the $V_{606}$ image to match the ground-based seeing 
(FWHM=1\arcsec).  Comparing the total $V_{606}$ flux in the aperture ($1.15\pm0.01\times10^{-15}$ \ergscm) to the 
reported \heii\ and \civ\ line fluxes ($4.07\pm0.04\times10^{-17}$ and $4.17\pm0.04\times10^{-17}$ \ergscm, respectively; Paper~I), 
we find that the \heii\ and \civ\ emission lines together contribute a small fraction of the $V_{606}$-band 
flux within this aperture ($\lesssim7$\%), 
and therefore we do not apply a correction to the $V_{606}$ image.  

\subsection{HST NICMOS Data}

Using the NICMOS NIC2 camera on HST, we obtained high-resolution imaging of the source in 
the $F110W$ ($J_{110}$) and $F160W$ ($H_{160}$) filters --- filters which at \zblob\ 
bracket the Balmer/4000\AA\ break.  The observations were taken during UT 2006 March 25 and 31, 
using a NIC-SPIRAL-DITH spiral dither pattern (3 point pattern with 0.6375\arcsec\ point spacing).\footnote{HST Cycle 14; GO 10591}   
The data were reduced primarily using NICRED \citep{magee07}.  Once the data were calibrated 
and corrected for electronic ghosts, pedestal, cosmic ray persistence, and count-rate non-linearity, 
we made the final image mosaics using MultiDrizzle \citep{jedrz05} and a set of custom bad pixel masks.  
The final images were supersampled to match the ACS pixel scale (0.05\arcsec/pixel) and have 
a field-of-view of 20\arcsec$\times$20\arcsec.  
The PSF sizes are FWHM=0.09\arcsec\ and 0.13\arcsec, as measured using the TinyTim PSF emulator, 
and the 5$\sigma$ point source limiting magnitudes are 27.2 and 27.1 mag (\apsize\arcsec\ diameter aperture) 
for the $J_{110}$ and $H_{160}$ imaging, respectively.

\subsection{Subaru Suprime-Cam Data}

In this work, we make use of deep Subaru \lya\ imaging that was obtained previously by \citet{pres08} using the
Subaru telescope and the SuprimeCam wide-field imager \citep{miya02}.  These observations used
an intermediate-band filter \ib\ ($\lambda_{c}\approx$4458\AA, $\Delta\lambda_{FWHM}\approx$201\AA),
centered on the \lya\ line at the redshift of the nebula.  The limiting magnitude of the \lya\ image
is 26.6 mag (5$\sigma$ in a 2\arcsec\ diameter aperture).
Additional details on the observations and data reduction can be found in \citet{pres08}.

\begin{figure}
\center
\includegraphics[angle=0,width=5in]{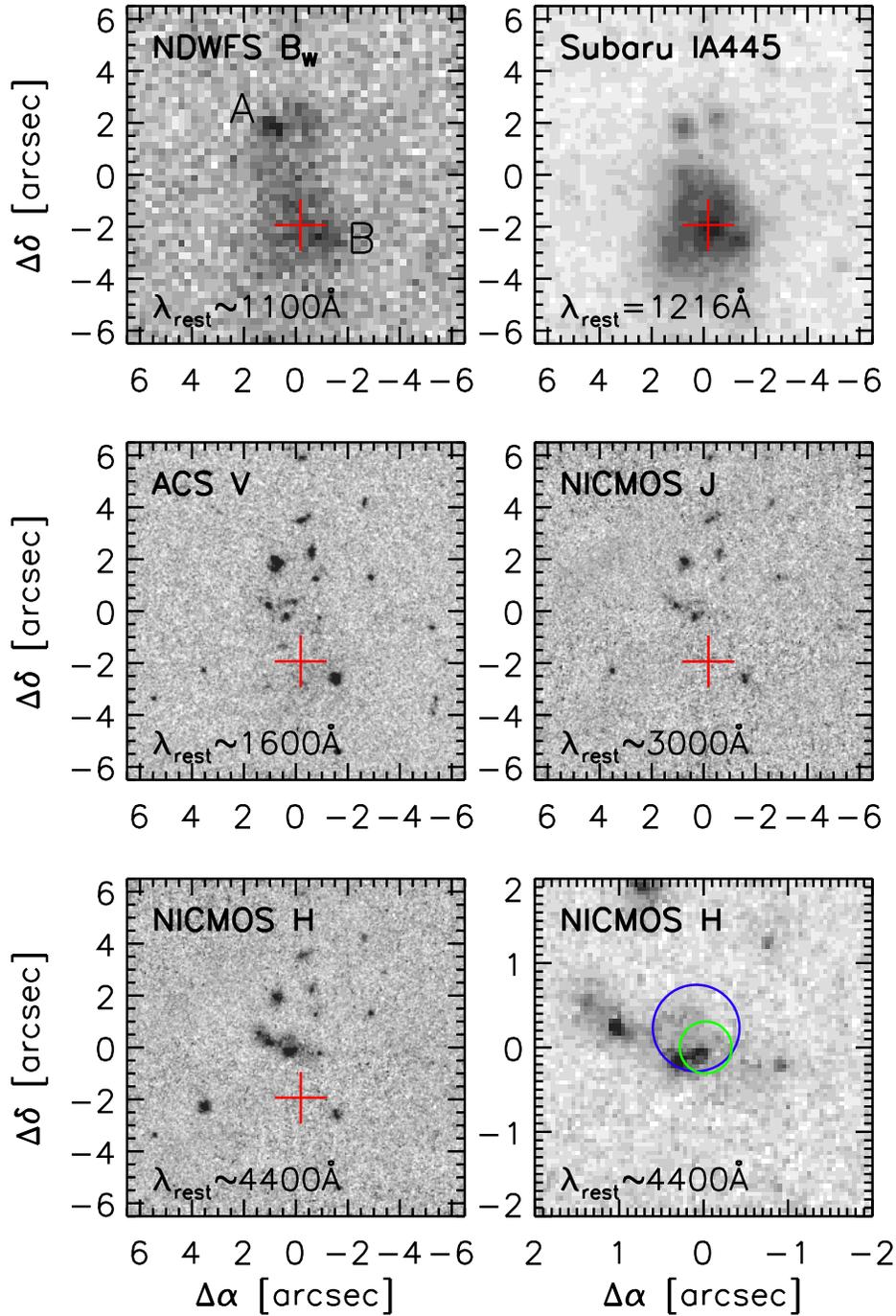}
\caption[Multi-wavelength Imaging of a \lya\ Nebula at $z\approx2.7$]
{Multi-wavelength imaging of LABd05.  In each panel, [0\arcsec,0\arcsec]
is centered on the location of the obscured AGN (see Section~\ref{sec:agn}), 
and the position of the \lya\ centroid measured from the Subaru imaging is shown (red cross).  
The compact components identified in ground-based imaging (Galaxies ``A" and ``B" from
Paper~I) are labeled on the NDWFS \bw\ image.  The lower right-hand panel 
shows a zoomed-in version of the NICMOS $H_{160}$ image with the position 
of the MIPS~24\micron\ and IRAC~3.6\micron\ source positions indicated 
\citep[blue and green $1\sigma$ error circles, respectively;][]{dey05,gor08}.  
At $z\approx2.656$, 1\arcsec\ corresponds to a physical scale of 7.96~kpc.
}
\label{fig:imall}
\end{figure}

We generated a Subaru line-only \lya\ image by subtracting off a smoothed version of
the ACS $V_{606}$-band image.  From the ground-based spectroscopy, we know that the source labeled 
``A" in the NDWFS \bw\ image (Figure~\ref{fig:imall}) is a Lyman Break galaxy (LBG) at the 
redshift of the system and shows little if any \lya\ emission or absorption (Paper~I).
A correct continuum subtraction should therefore leave the LBG with zero flux in the Subaru line-only \lya\ image.
We smoothed the $V_{606}$ image to match the PSF of the \lya\ image (FWHM=0.7\arcsec), resampled
to the same pixel scale as the \lya\ image, measured the flux of the LBG in both
the \lya\ and $V_{606}$-band image (using a 1.0\arcsec\ diameter aperture), scaled
the $V_{606}$ to match the flux of the LBG in the \lya\ image, 
and subtracted the two to create a continuum-subtracted \lya\ image.
Since the $V_{606}$ is a rather crude approximation to the continuum
in the \lya\ image, the accuracy of this continuum subtraction will vary with source color.
We note that for all the galaxies at the redshift of the system, the
subtraction should be relatively accurate; those that show residual emission in the \lya\ image
are likely \lya-emitters themselves.  On the other hand, in the case of a known interloper galaxy
at $z\approx3.2$ (labeled `B' in Figure~\ref{fig:imall}; Paper~I), 
we expect our continuum subtraction procedure to overestimate the
continuum (since at this redshift the \ib-band is sampling the continuum shortward of \lya\
which is typically depressed due to the \lya\ forest).
This is consistent with the slight evidence of oversubtraction that we see at the position 
of the interloper galaxy (Figure~\ref{fig:acslabel}).

\subsection{Image Registration}

To ensure accurate image registration, we generated catalogs of source
positions in each image using SExtractor \citep{ber96} and registered all images to the NDWFS \bw\
frame using the IRAF tasks {\it ccmap} and {\it ccsetwcs}.  We carried out the registration in three steps.
First, the ACS $V_{606}$ was convolved with a Gaussian kernel to match the PSF of the \bw\ image (FWHM=0.98\arcsec).  
We registered the 
smoothed ACS $V_{606}$ image to the NDWFS \bw\ image, applied the solution to the unsmoothed
ACS $V_{606}$ image, and then registered the 
ACS \heii, NICMOS $J_{110}$, and NICMOS $H_{160}$ images
to the unsmoothed ACS $V_{606}$ image.  Finally, we registered the ACS \lya\ image to the
ACS \heii\ image and the Subaru \ib\ image to the NDWFS \bw\ image.  
This sequential procedure was used to maximize the number of common sources
available to compute the astrometric solution for each image pair and to avoid compounding
registration errors.  
Table~\ref{tab:hstreg} details the number of sources used in the registration
and the final estimated astrometric uncertainty relative to the NDWFS \bw\ image.  
The NDWFS astrometry is tied to a frame defined by stars in the USNO-A2.0 catalog.  

\section{The Components of the Nebula}
\label{sec:components}
Our multi-wavelength observations show that LABd05 contains a number of compact galaxies, diffuse rest-frame 
UV continuum emission, smooth \lya\ emission, spatially-extended \heii\ emission, and an obscured AGN 
- all within a $\approx$10\arcsec\ region.  In this section, we explore each component of the \lya\ nebula 
system in detail.  In Section~\ref{sec:discussion}, we summarize the key morphological characteristics of this system 
and discuss the implications of these results for our understanding of the physical mechanisms at work in 
\lya\ nebulae, their evolutionary state, and the properties of the galaxies forming within them.

\subsection{Compact Sources}  

To determine the relative positions of all the sources in the vicinity of the \lya\ nebula 
we created a composite stack of the $V_{606}$, $J_{110}$, and $H_{160}$ images, after 
convolving the $V_{606}$ and $J_{110}$-band images to the same PSF as the $H_{160}$ 
image and dividing each image by the variance of the sky.  
We generated an initial list of source positions in the image stack using SExtractor 
\citep[3$\sigma$ threshold, minimum contiguous area of 4 pixels;][]{ber96}.   
All sources within 7\arcsec\ that are detected above the $5\sigma$ limiting magnitude in 
the $V_{606}$ band are labeled with ID numbers in Figure~\ref{fig:acslabel}.  
We then generated an additional catalog of sources positions using the unconvolved 
$V_{606}$-band image and the same parameters.  This ``$V_{606}$-only" catalog is necessary for our 
analysis of the number counts in the vicinity of LABd05 and the associated 
completeness corrections (Sections~\ref{sec:phot} and \ref{sec:overdense}). 
In four cases (\#1, 21, 26, and 46) where SExtractor did not deblend an apparent 
close object pair, we manually added a second source position to the catalog (\#57, 58, 59, 
and 60, respectively), as described in Section~\ref{sec:morph}.  
The current data are not sufficient to distinguish whether these objects are true companions 
or just morphological peculiarities (i.e., tidal features, dust lanes, etc.) associated with 
the primary object.  We choose to treat these pairs as separate objects; however, combining them does not 
significantly change our conclusions.  Individual postage stamps extracted from the image stack as well as 
the $V_{606}$, $J_{110}$, and $H_{160}$-band imaging are shown 
for all sources in Appendix~\ref{appendixA}.
 
\subsubsection{Optical and Near-Infrared Photometry}
\label{sec:phot}

We measured aperture photometry (\apsize\arcsec\ diameter apertures) in all three bands using the 
original unconvolved images and the positions derived from the image stack.  
The aperture size was chosen in order to contain as much flux as possible 
while minimizing contamination from neighboring sources.  
Aperture corrections of \apcorr\ were computed using the TinyTim PSF emulator and 
applied to the $V_{606}$, $J_{110}$, and $H_{160}$ photometry, respectively. 
The resulting photometry is given in Table~\ref{tab:galfitcompact}.  

Without knowing the intrinsic colors of sources as a function of magnitude, it is difficult to estimate the completeness of 
the stacked catalog.  Instead, we measured the completeness in the $V_{606}$-band alone using the standard approach.  
First, we generated simulated galaxies, modeled as Gaussian profiles with 
$R_{e}=0.12$\arcsec\ (1~kpc, the typical size of non-member sources in the field; Section~\ref{sec:morph}).  
We then inserted the simulated galaxies into the $V_{606}$-band imaging.  We generated a ``$V_{606}$-only" source catalog 
in the same manner as above and computed the completeness of our approach as a function of input $V_{606}$ magnitude.  
The 80\% and 50\% completeness limits are 27.6 and 28.0 mag in the $V_{606}$ band, respectively.

\subsubsection{The Obscured AGN}
\label{sec:agn}
Paper~I postulated the existence of an obscured AGN at the position of the MIPS source based on the strong 
24\micron\ emission and the shape of the full SED.  The corresponding detections in the IRAC bands agreed with the MIPS source 
position to within the astrometric uncertainty ($\approx$0\farcs5) and showed that the source had a power-law SED 
in the mid-infrared, characteristic of an obscured AGN \citep[e.g.,][]{alo06}.  Later IRS spectroscopy 
demonstrated that the infrared source is at the redshift of the \lya\ nebula 
\citep{colbert11}, and millimeter and submillimeter observations have confirmed that 
the full SED is best approximated by a Mrk~231 (i.e., AGN-dominated) template \citep[][]{bussmann09,yang12}.  
In Paper~I, the physical location of the AGN within the system was uncertain due 
to the lower resolution of the IRAC and MIPS imaging, 
but the centroid of the mid-infrared emission appeared to be offset to the north of the 
brightest \lya\ emission.  The addition of the HST/NICMOS imaging revealed an extremely red source 
(\#\agnid) located at the centroid of the MIPS 24\micron\ emission (Figure~\ref{fig:imall}) 
that is very centrally concentrated (Section~\ref{sec:morph}).  
This source shows a strong Balmer/4000\AA\ break --- it is barely detected in $V_{606}$ and $J_{110}$ 
but very bright in $H_{160}$ --- and is one of the \numgalagn\ sources flagged as members 
of the system (Section~\ref{sec:membership}).  
It is located in a crowded region, with 5 close neighbors within $\approx$1.5\arcsec\ ($\approx$12 projected kpc), 
and diffuse emission visible in the NICMOS $H_{160}$ band, suggestive of an ongoing merger.  
Since it is plausible to assume that the AGN lies near the deepest part of the gravitational potential well 
of this system, we will take the position of source \#\agnid\ as the center of the system for our subsequent analysis.  
The measured projected offset between the AGN and the centroid of the \lya\ emission 
(Section~\ref{sec:lya}) is $\approx$1.9\arcsec\ ($\approx$15~kpc).  

\subsubsection{Assessing System Membership}
\label{sec:membership}

Only two compact sources were previously identified from ground-based imaging (Paper~I; see NDWFS 
\bw\ image in Figure~\ref{fig:imall}).  
At the Northeast corner of the system is a compact source --- labeled ``Galaxy A" in Paper~I --- 
that ground-based spectroscopic follow-up showed to be an LBG at the redshift of the 
system.  The source at the Southwest corner --- named ``Galaxy B" in Paper~I --- was argued 
to be an interloping system based on the identification of \lya\ at 
$z\approx3.2$ in the ground-based spectrum.  The high resolution HST imaging resolves both 
of these objects into two components: Galaxy A is associated with objects 
\#\lbgidA\ and \lbgidB\ and Galaxy B contains knots \#\interidA\ and \interidB\ (see Section~\ref{sec:morph} and Appendix~\ref{appendixA}).  
Since the spectroscopic identification was done using ground-based spectroscopy that was unable 
to resolve the two components in each case, it is possible that these pairs are in fact due to chance coincidence.  
However, with separations of only $\sim0.2\arcsec$ (i.e., 1.6 kpc), chance projection is extremely unlikely, and 
the photometry shows that the colors of both components in each pair are similar.  
Given the ground-based spectroscopic redshift, the very small likelihood of a chance coincidence, and the similar colors, 
we will assume from here on that \#\lbgidA\ and \lbgidB\ are both associated with the \lya\ 
nebula system at $z\approx2.7$ and that \#\interidA\ and \interidB\ are both interlopers at $z\approx3.2$.   

\begin{figure}
\center
\includegraphics[angle=0,width=4in]{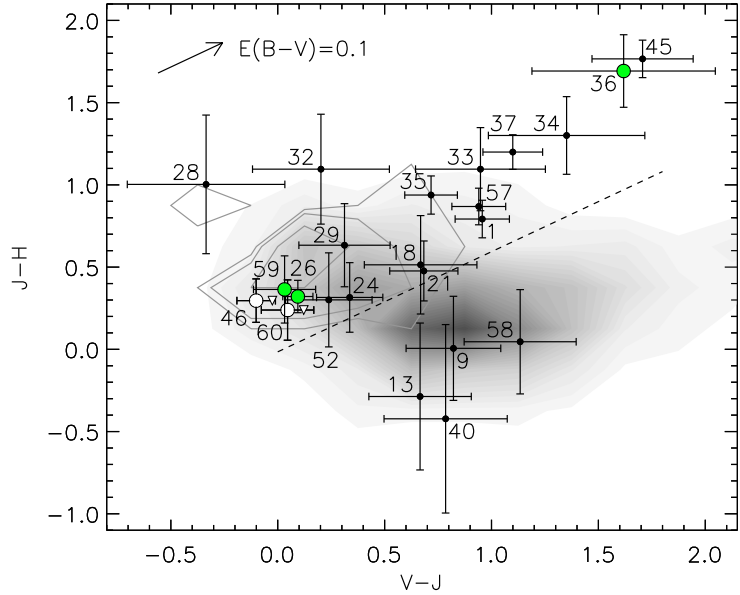}
\caption[$J-H$ vs. $V-J$ color of compact sources.]
{$J_{110}-H_{160}$ vs. $V_{606}-J_{110}$ color-color plot for compact sources in the vicinity of LABd05.  
Known spectroscopic members are denoted with large green circles (the LBG system, \#\lbgidA+\lbgidB, 
and the counterpart to the obscured AGN, \#\agnid).  
The known spectroscopic interlopers (\#\interidA+\interidB) are shown as measured (large open circles) 
and after correcting for the contribution of \lya\ emission to the $V_{606}$-band (open triangles; 
see Table~\ref{tab:galfitcompact}).  
The greyscale contours represents all galaxies from the HUDF above the magnitude limits of 
our $V_{606}$, $J_{110}$, and $H_{160}$ data; the line contours represent the subset 
with photometric redshifts at the redshift of LABd05 \citep[$z_{phot}=2.656\pm0.15$;][]{coe06}.  
The dashed black line indicates the division used for membership assignment (Section~\ref{sec:membership}).  
The appropriate reddening vector is shown for $E(B-V)=0.1$ computed at $z\approx2.656$.  
}
\label{fig:VJHcolor1}
\end{figure}

To determine the membership of the remaining sources, we make use of the measured optical/NIR colors.  
Since the NICMOS $J_{110}$ and $H_{160}$ bands straddle the Balmer/4000\AA\ break at the redshift of 
the nebula, sources within the system should show red $J_{110}-H_{160}$ colors if they 
have evolved enough with time.  While a full SED-fitting approach would be poorly constrained with only 
three bands, we can use this fact to identify other sources that are likely associated with the system.  
We start by selecting the sample of sources that are within a radius of 7\arcsec\ from the AGN (\#\agnid) and that 
are brighter than the $5\sigma$ limiting magnitude in all three bands.  In Figure~\ref{fig:VJHcolor1} 
we plot the $V_{606}-J_{110}$ versus $J_{110}-H_{160}$ colors of the resulting 
sample along with greyscale contours representing the expected color distribution of galaxies drawn from 
the field \citep[taken from the Hubble Ultra Deep Field, HUDF, which used the same instrument and filters as this work;][]{coe06}.  
The subset of HUDF galaxies with photometric redshifts consistent with the systemic redshift to within typical 
photometric redshift errors ($z_{photo}=2.656\pm0.15$) is shown with line contours.  
While a handful of sources have colors entirely consistent with being drawn from the field (i.e., they are near the 
peak of the HUDF greyscale), there is also a locus of objects that is broadly consistent with being at the systemic redshift 
(i.e., they are within the line contours) but that extends along a line roughly parallel to the reddening vector towards much redder colors.  
These red colors are quite unusual for typical field galaxies, a hint that this locus may be composed primarily 
of system members with varying amounts of dust. 

\begin{figure}
\center
\includegraphics[angle=0,width=4in]{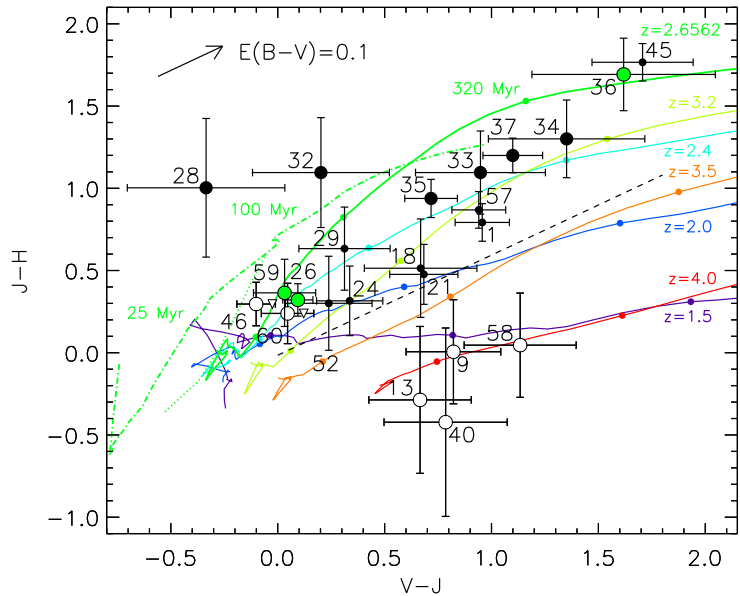}
\caption[$J-H$ vs. $V-J$ color of compact sources.]
{$J_{110}-H_{160}$ vs. $V_{606}-J_{110}$ color-color plot for compact sources in the vicinity of LABd05.  
Member sources in category ``M1" are shown as large filled circles --- with known spectroscopic members 
in green, i.e., the LBG system (\#\lbgidA+\lbgidB) and the obscured AGN (\#\agnid) --- and those in ``M2" 
as small filled circles.  Non-member sources (``NM") are shown as open circles; the known spectroscopic 
interlopers (\#\interidA+\interidB) are shown as measured (large open circles) and after correcting for 
the contribution of \lya\ emission to the $V_{606}$-band (open triangles; see Table~\ref{tab:galfitcompact}).  
A series of single stellar population model age tracks 
\citep[unreddened burst, solar metallicity, spanning burst ages of 5-1400~Myr;][]{bc03} 
are overplotted for different redshifts. 
The thick green line corresponds to $z\approx2.656$, the redshift of LABd05.  
The 25~Myr, 100~Myr, and 320~Myr positions along the tracks are indicated with small filled circles.  
For comparison, a constant star-forming model (dotted green line) and a low metallicity model 
($Z=0.0001$, dot-dashed green line) are shown, both spanning the same age range at the redshift of LABd05.  
The dashed black line indicates the division used for membership assignment (Section~\ref{sec:membership}).  
The appropriate reddening vector is shown for $E(B-V)=0.1$ computed at $z\approx2.656$.  
}
\label{fig:VJHcolor2}
\end{figure}

\begin{figure}
\center
\includegraphics[angle=0,width=4in]{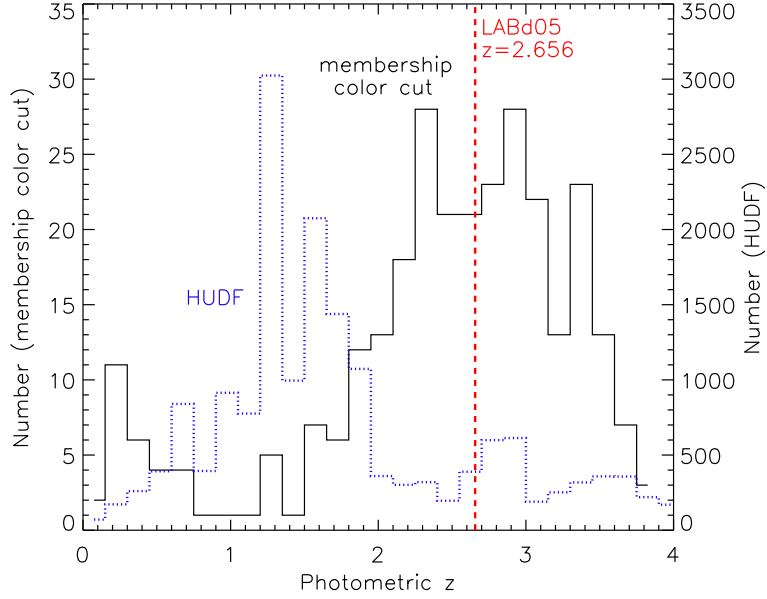}
\caption[Photometric redshift distribution]
{Photometric redshift distribution for galaxies in the HUDF that satisfy the 
color cut used for membership assignment in this work (black solid histogram) in comparison with the 
distribution for the full HUDF galaxy sample \citep[blue dotted histogram;][]{coe06}.  
This comparison demonstrates that the proposed color cut is successful at selecting high redshift sources 
(90\% at $z>1$) with the peak of the resulting redshift distribution centered on the redshift 
of LABd05 (red dashed line). 
}
\label{fig:photz}
\end{figure}

In Figure~\ref{fig:VJHcolor2}, we plot the same sample alongside 
a series of age tracks for simple stellar population models (single unreddened bursts, solar metallicity) 
at $z=1.5-4.0$ as well as a constant star-forming model (solar metallicity) and a low metallicity model ($Z=0.0001$), 
both at the systemic redshift \citep{bc03}.  
We again see that the locus discussed above consists of sources that are consistent with the systemic redshift 
if we allow for a low to moderate amount of dust extinction ($E(B-V)\approx0.0-0.4$ mag).  We draw a dividing line 
in color-color space with a slope parallel to the reddening vector in order to select sources that are 
consistent with the systemic redshift, given typical photometric errors.  Those to the upper left of the line 
we consider ``members" and those to the lower right, ``non-members" (designated ``NM").  As a check on the effectiveness 
of our approach, we apply the same color cut to the HUDF galaxy catalog and plot a histogram of the photometric redshifts 
for this sub-sample (Figure~\ref{fig:photz}).  The color cut is effective at selecting high redshift galaxies (90\% 
are at $z_{photo}\gtrsim1$), and in contrast to the redshift distribution for the full HUDF sample, the photometric 
redshifts of HUDF galaxies selected using this simple color cut 
are peaked at the systemic redshift of LABd05 ($z_{photo}\approx2.66$).  
On the other hand, our color cut is clearly approximate.  
The comparison sample selected from the HUDF contains a small subset (10\%) of galaxies with $z_{photo}\lesssim1$, 
which is consistent with the fact that for young ($\approx25$~Myr) single burst models the predicted 
colors at $z\lesssim1$ overlap those for higher redshifts.  
The color cut is also not able to reject 
the known interloper system at $z\approx3.2$ (\#\interidA\ and \#\interidB, or Galaxy B).  
From Figure~\ref{fig:photz}, this is not at all 
surprising, as the peak of the $z_{photo}$ distribution is broad, spanning $2.1\lesssim z_{photo}\lesssim3.2$.  
Furthermore, the fact that the interloper system is located at the young end of the age tracks in color-color space 
(Figure~\ref{fig:VJHcolor2}) is consistent with the detection of \lya\ emission (Paper~I).  

\begin{figure}
\center
\includegraphics[angle=0,width=6in]{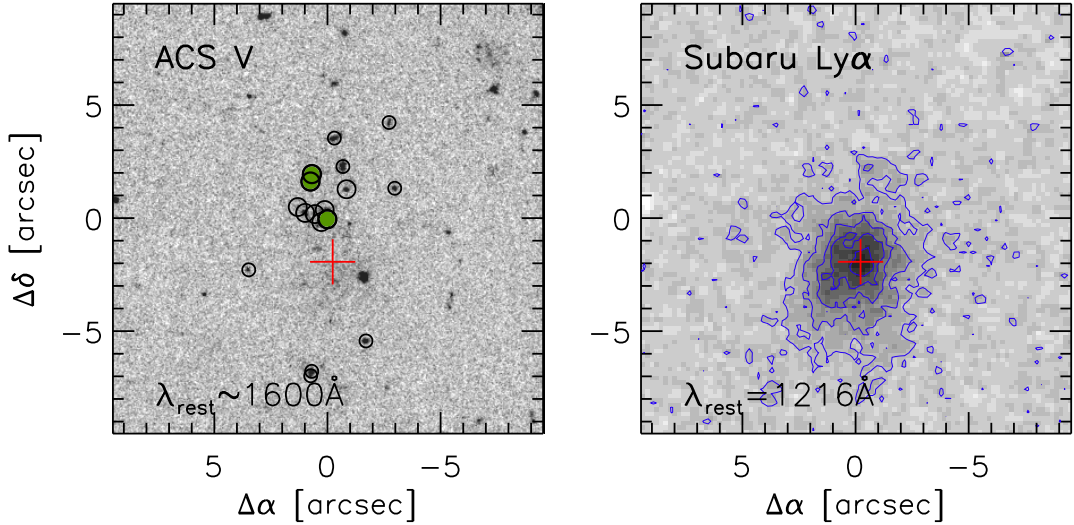}
\caption[Galaxy Members]
{LABd05 galaxy membership.  
The obscured AGN is centered at [0\arcsec,0\arcsec] in both panels, and the position of the \lya\ centroid 
measured from the Subaru imaging is shown as a red cross.  
{\bf Left:} The ACS $V_{606}$ image, with circles representing sources that have been 
flagged as members of the system based on optical/NIR colors and distance from the AGN 
(``M1'' spectroscopic members: filled green circles, ``M1'': large black circles, ``M2'': small black circles; see Section~\ref{sec:membership}).  
{\bf Right:} The continuum-subtracted Subaru \ib\ (\lya) imaging, with contours at 
\lya\ surface brightness levels of [1, 3, 5, 7, 9]$\times10^{-17}$ erg 
cm$^{-2}$ s$^{-1}$ arcsec$^{-2}$.  
Note that the \lya\ emission is offset by $\gtrsim$1.9\arcsec$\approx$15 projected kpc from all of the member galaxies.
}
\label{fig:immember}
\end{figure}

We assess the robustness of our membership assignment further in Section~\ref{sec:properties}.  
At this point, we subdivide the member sample based on the projected distance from the AGN (\#\agnid).  The 
\nummone\ sources within the radius of \lbgdist\arcsec\ (chosen to include all spectroscopically confirmed 
members, \#\lbgidA, \lbgidB, \agnid) are considered to be members with high confidence and designated ``M1".  
The \nummtwo\ members outside this radius are designated ``M2".  
In Figure~\ref{fig:immember}, we see that all the member sources (circled) 
are offset by $1.9-6.7$\arcsec\ ($15-53$ projected kpc) 
from the peak of the \lya\ emission (Section~\ref{sec:lya}), i.e., they effectively reside at the outskirts of the \lya\ nebula.

\subsubsection{Sizes and Morphologies}
\label{sec:morph}
Our next step was to use GALFIT \citep{peng02} to derive sizes and morphologies for the sources within the vicinity of LABd05.  
Visual inspection of the $V_{606}$ image revealed that in addition to the many compact galaxies in the region, there 
is diffuse, spatially extended emission.  In order to avoid biasing the fits for the compact sources, 
we began by subtracting off an approximate fit to this diffuse emission using GALFIT.  
We then fit all remaining compact sources within 7\arcsec\ of the AGN that have a peak surface brightness 
brighter than 25.3 mag arcsec$^{-2}$ and an isophotal magnitude brighter than 29.0 AB mag in the $V_{606}$ band.  
To speed up the process of fitting so many sources simultaneously, we performed 
the fits in small batches initially, i.e., fitting $1-7$ nearby sources at a time while masking the remaining sources.  
After experimentation showed that sources \#\specbotidA, \spectopidA, \lbgidA, and \interidA\ were not being 
fit well using single components, sources \#57, 58, 59, and 60, respectively, were manually added to the catalog.  
During the initial batch-fitting, the position of sources \#57, 58, and 59 were then masked, while \#60 was fit normally. 
Once appropriate fitting parameters had been determined for all sources, we performed a final run of GALFIT on the 
original image fixing all parameters for the compact sources to the pre-derived values.  We allowed the few remaining 
special cases (\#57, 58, and 59) as well as the diffuse continuum component to be fit freely by GALFIT.  
The entire fitting process was repeated using the fitting parameters derived for the diffuse continuum component 
in this final step as the initial guess for the first step.  
The resultant composite fit and corresponding residual images are shown in Figure~\ref{fig:galfitV}.  
The centroids and morphological results for the compact sources derived from the $V_{606}$ band are given in Table~\ref{tab:galfitcompact} 
and for the diffuse component in Table~\ref{tab:galfitdiffuse}.  
We discuss this diffuse UV continuum component in further detail in 
Section~\ref{sec:diffusecontinuum}.

\subsection{Diffuse Components}

In addition to the population of compact sources, LABd05 hosts several nearly coincident 
diffuse emission components.

\subsubsection{Diffuse \lya\ Emission}
\label{sec:lya}

\begin{figure}
\center
\includegraphics[angle=0,width=7in]{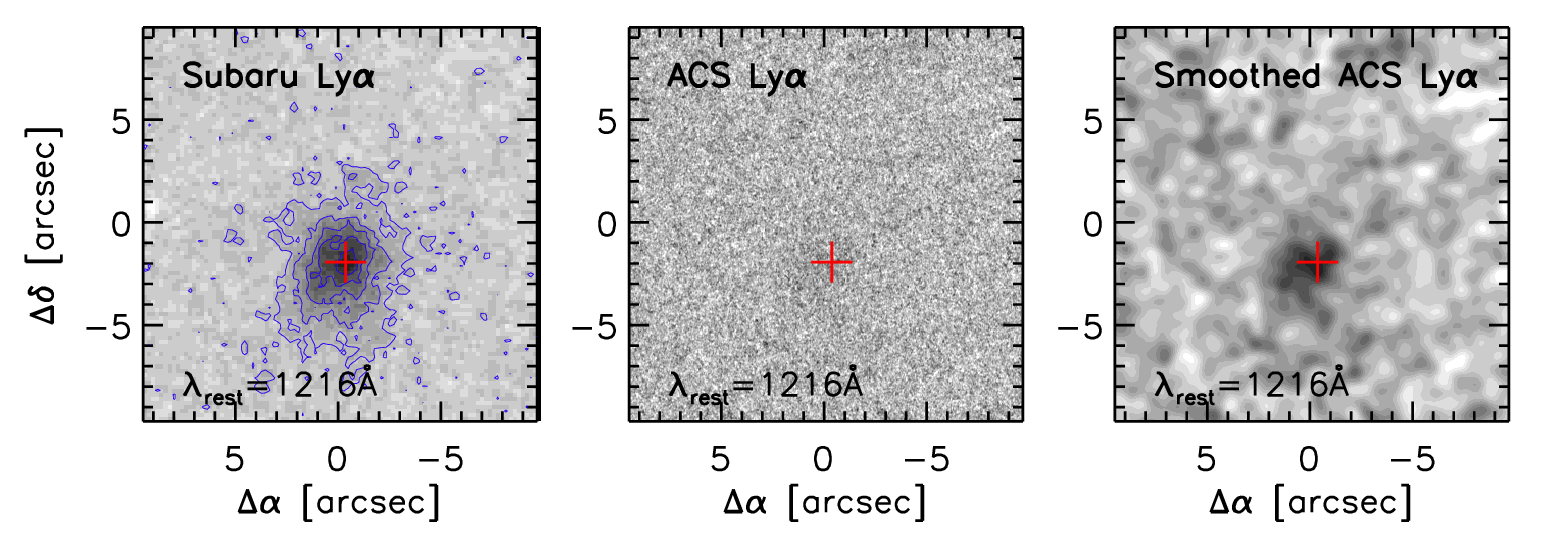}
\caption[\lya\ Imaging of LABd05]
{\lya\ imaging of LABd05.  
The obscured AGN is centered at [0\arcsec,0\arcsec] in all panels, and the position of the \lya\ centroid 
measured from the Subaru imaging is shown as a red cross.  
{\bf Left:} The continuum-subtracted Subaru \ib\ (\lya) imaging, with contours at 
\lya\ surface brightness levels of [1, 3, 5, 7, 9]$\times10^{-17}$ erg 
cm$^{-2}$ s$^{-1}$ arcsec$^{-2}$.  
{\bf Middle:} The ACS continuum-subtracted \lya\ image.  
{\bf Right:}  The ACS continuum-subtracted \lya\ image smoothed to match the PSF of 
the ground-based Subaru imaging (FWHM=0.7\arcsec).
No compact knots or clumps are detected in the ACS \lya\ imaging.
}
\label{fig:imlya}
\end{figure}

If the \lya\ nebula is powered by sources within the nebula itself, we would expect that 
to be reflected in the morphology of the \lya\ emission.  We used the high resolution ACS \lya\ imaging to 
look for spatially resolved knots or clumps that could signal the locations of the ionizing sources for the nebula.  
However, we find that the morphology of 
the \lya\ nebula is smooth, with no significant substructure (Figure~\ref{fig:imlya}).  
At the depth of these observations we would have detected point-source regions down to a \lya\ 
flux of $F_{Ly\alpha}=2.6\times10^{-17}$ \ergscm\ (5$\sigma$ in \apsize\arcsec\ diameter apertures) or 
\lya\ luminosity of $L_{Ly\alpha}=1.5\times10^{42}$ \ergss.  
This corresponds to a star-formation rate of $\approx1.3$~$M_{\odot}$~yr$^{-1}$ 
\citep[assuming Case B, $L_{Ly\alpha}=8.7\times L_{H\alpha}$;][]{ken98review}.  
We can also rule out high surface brightness clumps within the cloud down to a peak surface brightness limit of 
$\approx$4.0$\times$10$^{-16}$ \ergscmarc (3$\sigma$).

Deeper \lya\ imaging at ground-based resolution obtained with the Subaru Telescope showed 
that the \lya\ emission on larger scales has a smooth elliptical morphology that is well 
fit by an exponential disk profile.
Using GALFIT, we derived a centroid, effective radius, S\'ersic index, position angle, 
and axis ratio for the \lya\ emission (Table~\ref{tab:galfitdiffuse}).  
The resulting surface brightness profile of the \lya\ emission is shown in Figures~\ref{fig:sbprofile1} and \ref{fig:sbprofile2}.  
Surface brightness profiles were also computed for both the Subaru \lya\ imaging and the ACS \lya\ imaging 
using the IRAF task {\it ellipse}, where the ellipse parameters were constrained to the centroid, single position angle, and axis ratio of 
the GALFIT parametric fit to the Subaru \lya\ imaging.  
In addition, we measured aperture photometry at the center of the nebula using a small aperture (1.6\arcsec\ diameter) 
chosen to minimize contamination from neighboring compact sources (Table~\ref{tab:galfitdiffuse}).  

\subsubsection{Diffuse He~II Emission}  
\label{sec:heii}

\begin{figure}
\center
\includegraphics[angle=0,width=4in]{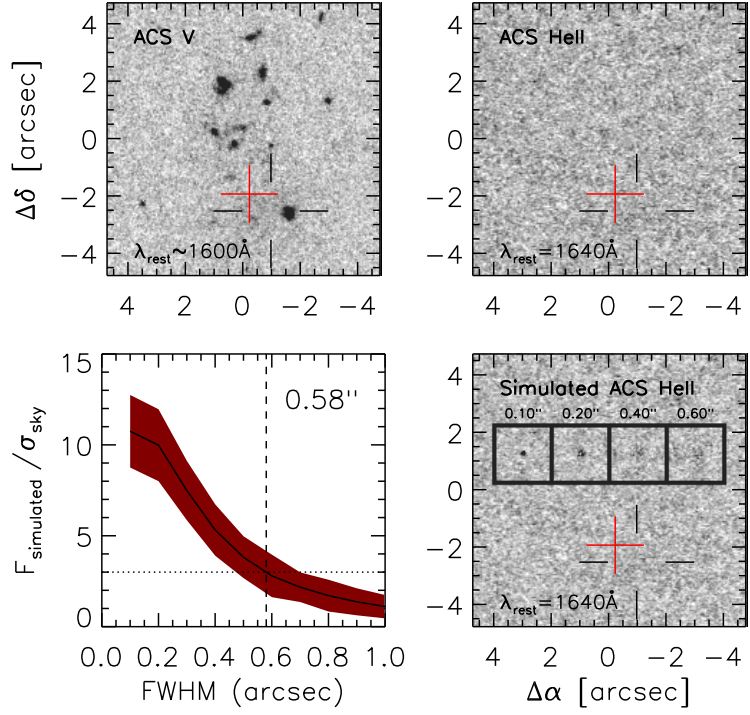}
\caption[\lya\ and \heii\ Imaging]
{Limits on the size of the \heii-emitting region derived from the continuum-subtracted ACS \heii\ imaging.  
In all panels, the position of the \lya\ centroid measured from the Subaru imaging is shown (red cross), 
and the approximate position of the \heii\ emission, as measured from ground-based spectroscopy, is indicated with black open crosshairs 
at [-0.9\arcsec,-2.5\arcsec].  
{\bf Top:}  ACS $V_{606}$ and continuum-subtracted \heii\ imaging.  
{\bf Bottom Left:}  
Ratio of the recovered flux of simulated \heii\ sources with varying sizes ($F_{simulated}$) 
divided by the $1\sigma$ limiting flux of the image ($\sigma_{sky}$), shown as a function of the FWHM of 
the simulated source (\apsize\arcsec\ diameter apertures).  
The red band denotes the full range of results from \numtrials\ Monte Carlo trials.  $F_{simulated}/\sigma_{sky}=3$ is shown (dotted line) along 
with the derived lower limit on the size of the \heii-emitting region (FWHM~$>\minheii\arcsec$; dashed line).  
Any source smaller than this would have been detected at $>3\sigma$ (Section~\ref{sec:heii}).  
{\bf Bottom Right:}  A simulated \heii\ image containing a representative set of model sources 
(FWHM~$=[0.1\arcsec,0.2\arcsec,0.4\arcsec,0.6\arcsec]$).
}
\label{fig:imlyaheii}
\end{figure}

From ground-based spectroscopy, LABd05 is known to have strong \heii\ and \civ\ emission (Paper~I) 
located near the center of the \lya\ emission ([-0.9\arcsec,-2.5\arcsec]; Figure~\ref{fig:imlyaheii}).  
The goal of our \heii\ imaging was to better localize the \heii-emitting region within the system, 
and given the depth of our observations, we would have detected the \heii\ emission at a $SNR\approx10$ 
if it were emitted as a point source.  
However, the ACS \heii\ imaging showed no detection down to 
a 5$\sigma$ point source limiting flux of $1.9\times10^{-17}$ \ergscm\ (\apsize\arcsec\ diameter aperture) 
and a 3$\sigma$ surface brightness limit of $3.3\times10^{-16}$ \ergscmarc\ (Figure~\ref{fig:imlyaheii}).  
While we cannot, therefore, pinpoint the location of the \heii\ emission beyond what was 
determined in Paper~I, this non-detection does put a constraint on the size of the \heii-emitting region.  
To quantify this, we inserted a series of simulated \heii\ sources, modeled as Gaussian 
profiles scaled to match the measured \heii\ 
flux and a range of FWHM sizes of 0.1$-$1.0\arcsec, into the \heii\ image 
(Figure~\ref{fig:imlyaheii}).  For each of \numtrials\ Monte Carlo trials, we 
measured the observed flux of the simulated source ($F_{simulated}$) as a function of FWHM and 
determined the FWHM for which the simulated source is detected at $F_{simulated}/\sigma_{sky}=3$, where 
$\sigma_{sky}$ is the $1\sigma$ limiting flux of the \heii\ image (\apsize\arcsec\ diameter aperture).  
We conclude that the source of \heii\ must be 
extended with a FWHM~$>\minheii\arcsec$ in order to be undetected in our \heii\ imaging.  
At the same time, the fact that the previous long-slit spectroscopy did not resolve the \heii\ line puts 
an upper limit of $\sim$1\arcsec\ on its true spatial extent (Paper~I).  
This size range corresponds to a \heii-emitting region that spans 
$\sim\minheiikpc-\maxheiikpc$~kpc at the redshift of the nebula.  
These constraints are summarized in Table~\ref{tab:galfitdiffuse}.

\subsubsection{Diffuse Continuum Emission}  
\label{sec:diffusecontinuum}

\begin{figure}
\center
\includegraphics[angle=0,width=4in]{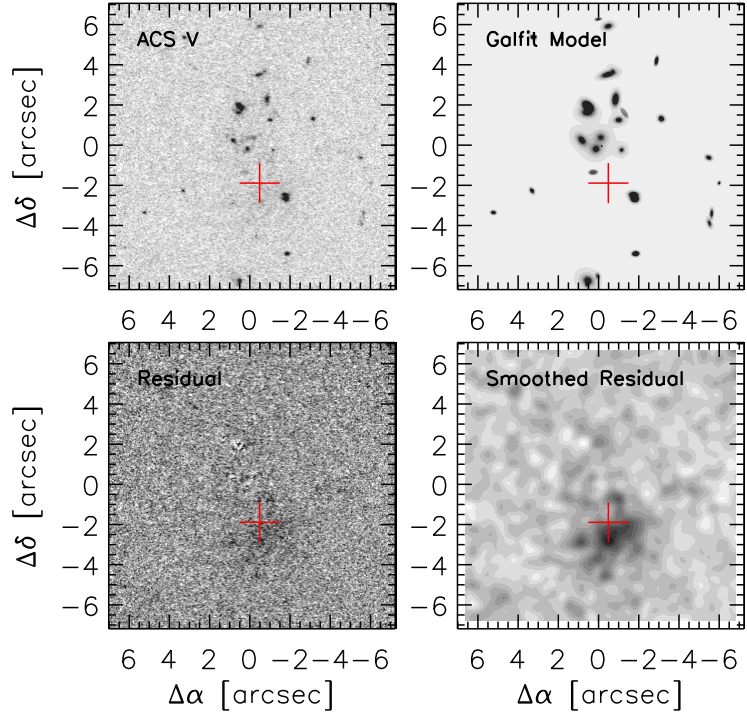}
\caption[GALFIT Fit to $V_{606}$ Data.]
{GALFIT parametric fit to the $V_{606}$ imaging of LABd05.  {\bf Top:} The original $V_{606}$ image and GALFIT model.  
{\bf Bottom:}  The residual image and smoothed residual image (FWHM~$=10$~pix~$=0.5$\arcsec\ Gaussian 
kernel) reveal a diffuse UV continuum component.  
Note that the diffuse $V_{606}$ continum emission is nearly coincident with the \lya\ centroid (red cross; an offset of 
$\approx\lyauvoffset$\arcsec), and that both components are in turn offset by $\approx$1.9-2.6\arcsec$\approx$15-21 
projected kpc from the position of the obscured AGN at [0\arcsec,0\arcsec].
}
\label{fig:galfitV}
\end{figure}

The broad-band ACS $V_{606}$ data revealed diffuse emission located near the 
center of the \lya\ nebula (Figure~\ref{fig:galfitV}; Section~\ref{sec:morph}).  
As \heii\ and \civ\ emission at the systemic 
redshift are located within the $V_{606}$ bandpass, and as the band also straddles 
\lya\ at the redshift of Galaxy B (the interloper system at $z\approx3.2$), we 
must first address the question of whether this diffuse $V_{606}$-band emission could 
be dominated by line emission.  
However, as we showed in Section~\ref{sec:acs}, the \heii\ and \civ\ emission lines can 
contribute at most 7\% of the $V_{606}$-band flux.  The \lya\ emission from Galaxy B 
($F_{Ly\alpha}=5.15\times10^{-17}$ \ergscm; Paper~I) can contribute at most another 
4\% of the $V_{606}$-band flux, but even this negligible fraction is likely a gross overestimate.  
The spectroscopic data shows that the \lya\ emission from Galaxy B is compact, centered on the 
interloper system, rather than spatially extended (Paper~I).  We conclude therefore that the 
diffuse $V_{606}$-band emission seen in LABd05 is indeed rest-frame UV continuum emission. 

As described in Section~\ref{sec:morph}, we measured the centroid, size, and luminosity of this component 
using GALFIT, avoiding contamination from nearby compact sources during the fitting process 
by constraining all the fitting parameters for the compact sources to the previously-derived values.  
Table~\ref{tab:galfitdiffuse} gives the position, magnitude, effective radius, S\'ersic index, position angle, 
and axis ratio of the diffuse component.  The diffuse continuum component is well-fit by an exponential disk profile and 
the centroid is nearly coincident with that of the \lya\ emission (offset by $\approx\lyauvoffset$\arcsec).  
We measured aperture photometry at the center of the nebula in all three bands using a small aperture (1.6\arcsec\ diameter) chosen to minimize 
contamination from neighboring compact sources (Table~\ref{tab:galfitdiffuse}).  
The surface brightness profile of the diffuse continuum emission was also computed using the 
IRAF task {\it ellipse}, where the ellipse parameters were constrained to the centroid, single position angle, and axis ratio of 
the GALFIT parametric fit (Figures~\ref{fig:sbprofile1} and \ref{fig:sbprofile2}).

At the depth of these observations we would have detected point-source regions with fluxes of 
$3.5\times$10$^{-17}$ \ergscm\ (5$\sigma$, \apsize\arcsec\ diameter aperture), corresponding to a 
stellar mass limit of $\approx3.1\times10^{7}$ M$_{\odot}$, 
assuming a 25~Myr simple stellar population \citep{bc03}.  
We can also rule out high surface brightness clumps down to a peak surface brightness limit of 
$\approx$9.5$\times$10$^{-16}$ \ergscmarc (3$\sigma$).  
However, it is worth remembering that if there are many lower mass sources distributed over this area, 
they would appear as unresolved diffuse continuum emission.

\section{Discussion}
\label{sec:discussion}

\subsection{The Substructure of LABd05}
\label{sec:substructure}

The high resolution imaging of HST provides a precise look at the sub-kiloparsec morphology 
of LABd05, with important implications for our understanding of the substructure 
of \lya\ nebulae and the properties of the galaxies forming within them, as well as for determining 
what power sources are ultimately responsible for the \lya\ emission.  
The key morphological characteristics of this \lya\ nebula system are the following: 
 
\begin{itemize}
\item Many Compact, Low Luminosity Galaxies: The system hosts \numgalagn\ primarily small, disky, low luminosity galaxies including an obscured AGN.   
\item Offset Morphology: All the compact sources within the system are located $\gtrsim$20~kpc away from the peak and centroid of the \lya\ nebula. 
\item No Central Galaxy: LABd05 has no central galaxy or compact source brighter than $\approx$0.03~$L^{*}$ visible within the highest surface brightness 
region of the \lya\ nebula.  
\item Diffuse Line and Continuum Emission: While the compact sources appear to avoid the region of the nebula, there are three nearly spatially coincident 
extended emission components located near the center of the nebula: \lya, \heii, and rest-frame UV continuum emission.  
\item Smooth, Non-filamentary Morphology: Both the \lya\ emission and the diffuse UV continuum emission are smooth 
and show surface brightness profiles that are consistent with exponential disks ($n\sim0.7-0.8$) with moderate axis 
ratios ($b/a\sim0.7-0.8$), i.e., they are not particularly clumpy or filamentary.
\item Similar UV and Ly$\alpha$ Surface Brightness Profiles: The \lya\ and UV continuum surface brightness profiles, while 
not identical, are comparable in shape and extent.
\end{itemize}

In what follows, we discuss each of the key morphological findings in detail.  Focusing first on the 
compact sources, we show that the LABd05 system is overdense relative to the field, 
hosting a population of small, disky, low luminosity galaxies that, while numerous, are irrelevant to the ionization 
of the nebula (Section~\ref{sec:compact}).  
We then compare the remaining observed morphological characteristics to the expectations for \lya\ nebulae 
driven by superwind outflow, cold flow, and resonant scattering scenarios and find significant discrepancies 
(Sections~\ref{sec:offset}$-$\ref{sec:sbprofile}).  We end with a discussion of the possibility that the LABd05 
system is a forming galaxy group (Section~\ref{sec:implications}).

\subsection{Many Compact, Low Luminosity Galaxies}
\label{sec:compact}

\subsubsection{An Overdense Region}
\label{sec:overdense}

\begin{figure}
\center
\includegraphics[angle=0,width=7in]{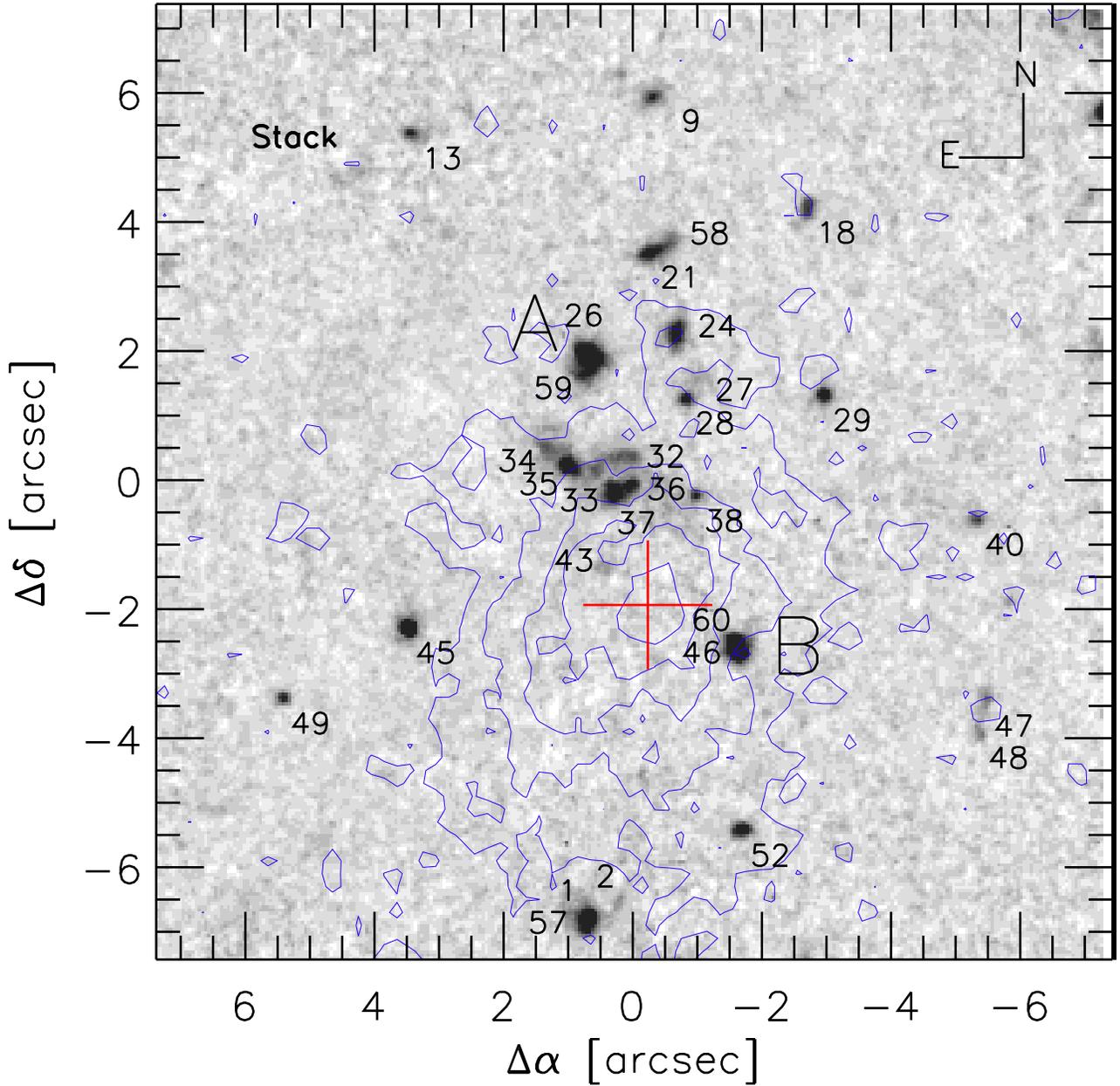}
\caption[Compact Sources within LABd05]
{A composite image of LABd05 made by stacking the $V_{606}$, $J_{110}$, and $H_{160}$ imaging.  
All compact sources detected above the $5\sigma$ limiting magnitude in the $V_{606}$ band and 
located within 7\arcsec\ of the obscured AGN (\#\agnid, located at [0\arcsec,0\arcsec]) are labeled with the ID 
number used in Table~\ref{tab:galfitcompact}.  
The contours correspond to \lya\ surface brightness levels of [1, 3, 5, 7, 9]$\times10^{-17}$ erg 
cm$^{-2}$ s$^{-1}$ arcsec$^{-2}$, as measured from the continuum-subtracted Subaru \ib\ (\lya) imaging.  
The position of the \lya\ centroid measured from the Subaru imaging is shown (red cross).  
}  
\label{fig:acslabel}
\end{figure}

\begin{figure}
\center
\includegraphics[angle=0,width=4in]{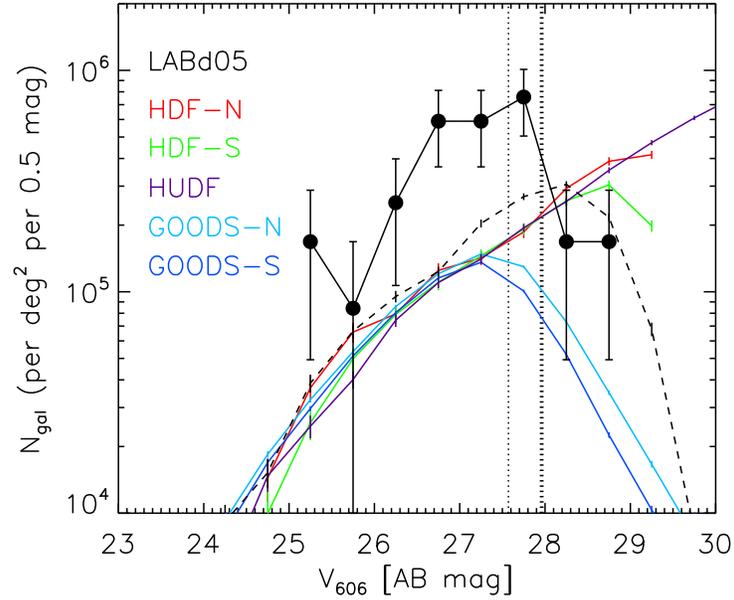}
\caption[Number Counts in LABd05]{
Number counts for sources detected above the $5\sigma$ limiting magnitude in the $V_{606}$ band that
are located within a 7\arcsec\ radius of the AGN in LABd05 (black solid line) and over the entire 
207\arcsec$\times$205\arcsec\ ACS pointing (dashed black line).   
The 80\% and 50\% completeness limits are indicated (thin and thick dotted lines, respectively).  
The $V_{606}$-band number counts from the HDF-N and HDF-S \citep[red and green lines;][]{williams96,caser00}, 
the HUDF \citep[purple line;][]{beck06,coe06}, and the GOODS-N and GOODS-S fields \citep[light and dark 
blue lines;][]{giav04} are shown, remeasured using aperture magnitudes consistent with our analysis. 
}
\label{fig:numbercounts}
\end{figure}

\begin{figure}
\center
\includegraphics[angle=0,width=4in]{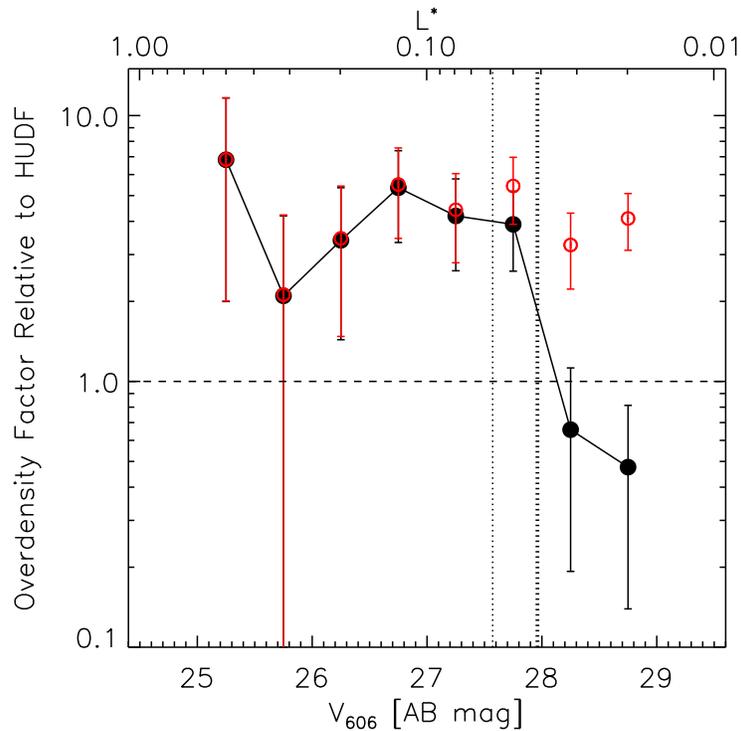}
\caption[Overdensity Factor for Galaxies Within a \lya\ Nebula.]
{Overdensity scale factor, relative to the field number counts from HUDF, of galaxies 
within a 7\arcsec\ radius of the AGN 
that are detected above the $5\sigma$ limiting 
magnitude in the $V_{606}$ band (black solid line).  
The LABd05 region is overdense by at least a factor of $\approx\overdenA$ relative to the field.  
The 80\% and 50\% completeness limits (thin and thick dotted lines, respectively) and the result of 
applying the completeness correction are shown (red open circles).  
The top axis is labeled in terms of $L^{*}$ at $z\approx3$ \citep[$M^{*}=-20.8$;][]{red08}.  
}
\label{fig:lumfuncscale}
\end{figure}

Even without determining the membership of individual nearby galaxies, it is clear 
from the high resolution ACS and NICMOS imaging that there are a large number of compact 
sources in the vicinity of LABd05, more than would be expected for a region of this size in the field.  
Figure~\ref{fig:numbercounts} shows the $V_{606}$-band number counts for sources detected above the 
$5\sigma$ limiting magnitude and located within a 7\arcsec\ radius of the AGN in LABd05.  
For comparison we show number count measurements drawn from our entire $207\arcsec\times205\arcsec$ ACS 
pointing as well as field measurements taken from the Hubble Deep Field-North and -South 
\citep[HDF-N, HDF-S;][]{williams96,caser00}, the HUDF \citep{beck06,coe06}, and the GOODS-North and -South fields \citep{giav04}.  
Dividing the LABd05 number counts by those from the HUDF 
shows that the \lya\ nebula system is overdense by at least factor of $\overdenA\pm\overdensigA$ relative to 
the field at magnitudes of $V_{606}<27.5$ (Figure~\ref{fig:lumfuncscale}), with the uncertainty computed 
assuming Poisson statistics.  If we further restrict the region of interest to a 2.5\arcsec\ radius region 
encompassing just the group of objects lying immediately to the north of the nebula 
(i.e., centered at [0\arcsec,+1.5\arcsec] in Figure~\ref{fig:acslabel}), 
the overdensity is a factor of $\overdenC\pm\overdensigC$ above the field.  
We can derive a better estimate of the true overdensity factor relative to the field 
at this redshift by applying the membership color cut discussed in Section~\ref{sec:membership} 
to both the LABd05 and HUDF galaxy catalogs.  In this case, the overdensity factor of the 7\arcsec\ radius 
region approaches $\overdenB\pm\overdensigB$ at magnitudes of $V_{606}<27.5$.  
These high overdensity factors are not surprising if LABd05 is indeed a region of active galaxy 
formation destined to become a galaxy group or cluster.

\subsubsection{Properties of the Member Galaxies}
\label{sec:properties}

\begin{figure}
\center
\includegraphics[angle=0,width=4in]{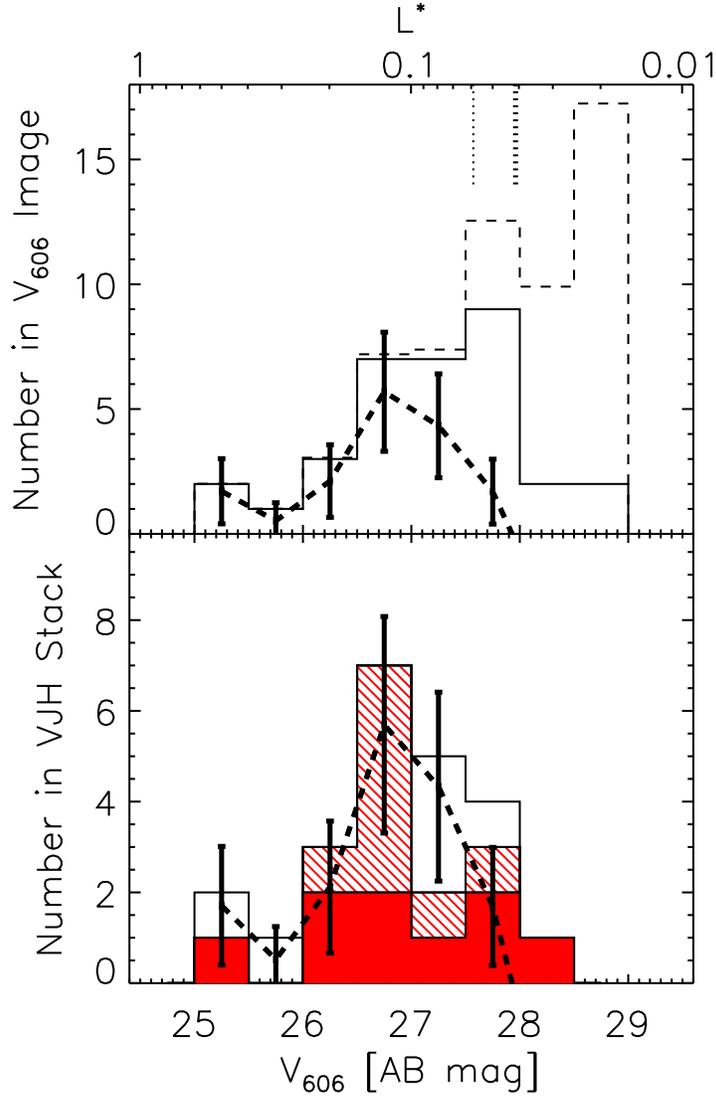}
\caption[Luminosity Function of Galaxies Within a \lya\ Nebula.]
{Luminosity function of galaxies within LABd05.  
{\bf Top:}  Sources detected above the $5\sigma$ limiting magnitude 
in the $V_{606}$ band and located within a 7\arcsec\ radius of the AGN (black open histogram) 
and corrected for incompleteness (thin dashed histogram).  The 80\% and 50\% completeness limits 
are shown (thin and thick vertical dotted lines, respectively).  
{\bf Bottom:}  Sources detected above the $5\sigma$ limiting magnitude 
{\it in all three bands} and located within a 7\arcsec\ radius of the AGN (black open histogram).  
The filled and hatched histograms represent galaxies in groupings ``M1'' and ``M2'', respectively (Section~\ref{sec:membership}).  
The thick dashed curve in both panels is a ``statistical'' luminosity function for LABd05, as described 
in Section~\ref{sec:properties}.  
The top axis is labeled in terms of $L^{*}$ at $z\approx3$ \citep[$M^{*}=-20.8$;][]{red08}.
}
\label{fig:lumfunc}
\end{figure}

The previous section looked at the question of whether this region is overdense in a statistical sense relative 
to the field, i.e., without drawing on any knowledge of system membership.  
In Figure~\ref{fig:lumfunc}, we again show the $V_{606}$ number counts (top panel), but 
in addition, we make use of our membership assignment to plot the number counts for sources 
that are likely within the system (bottom panel), i.e., the observed luminosity function.  
We find that the observed luminosity function of the M1+M2 subset is largely 
consistent with what we would predict statistically by taking the observed $V_{606}$ number counts for the 
LABd05 region and simply subtracting off the expected $V_{606}$ number counts for the field, as derived from the HUDF 
(Figure~\ref{fig:lumfunc}, dashed line).
This good agreement between the statistical treatment of the previous section and the 
individual object membership treatment gives us confidence both that our membership assignment based on optical/NIR colors 
is reasonable and that our estimate of the observed luminosity function of the system is relatively robust even if our 
membership assignment is not perfect on an object-by-object basis.  
To our knowledge, this is the first time a luminosity function has been estimated for an individual 
Ly$\alpha$ nebula system, and it implies that LABd05 is dominated by low luminosity galaxies.  
Assuming that $M^{*}$ at $z\approx2.7$ is $\approx-20.8$ \citep[AB mag at rest-frame 1700\AA;][]{red08}, which 
corresponds to $m^{*}\approx24.49$ AB in apparent $V_{606}$ 
magnitudes,\footnote{The central wavelength of $V_{606}$ is 
$\lambda_{c}\approx5907$\AA\ observed $\approx1600$\AA\ in the rest frame at $z\approx2.7$} 
we find that the LABd05 system is dominated by $\approx0.1 L^{*}$ galaxies.  
Only one galaxy, the LBG (\#\lbgidA), is more luminous than $\approx0.4 L^{*}$.  

\begin{figure}
\center
\includegraphics[angle=0,width=4in]{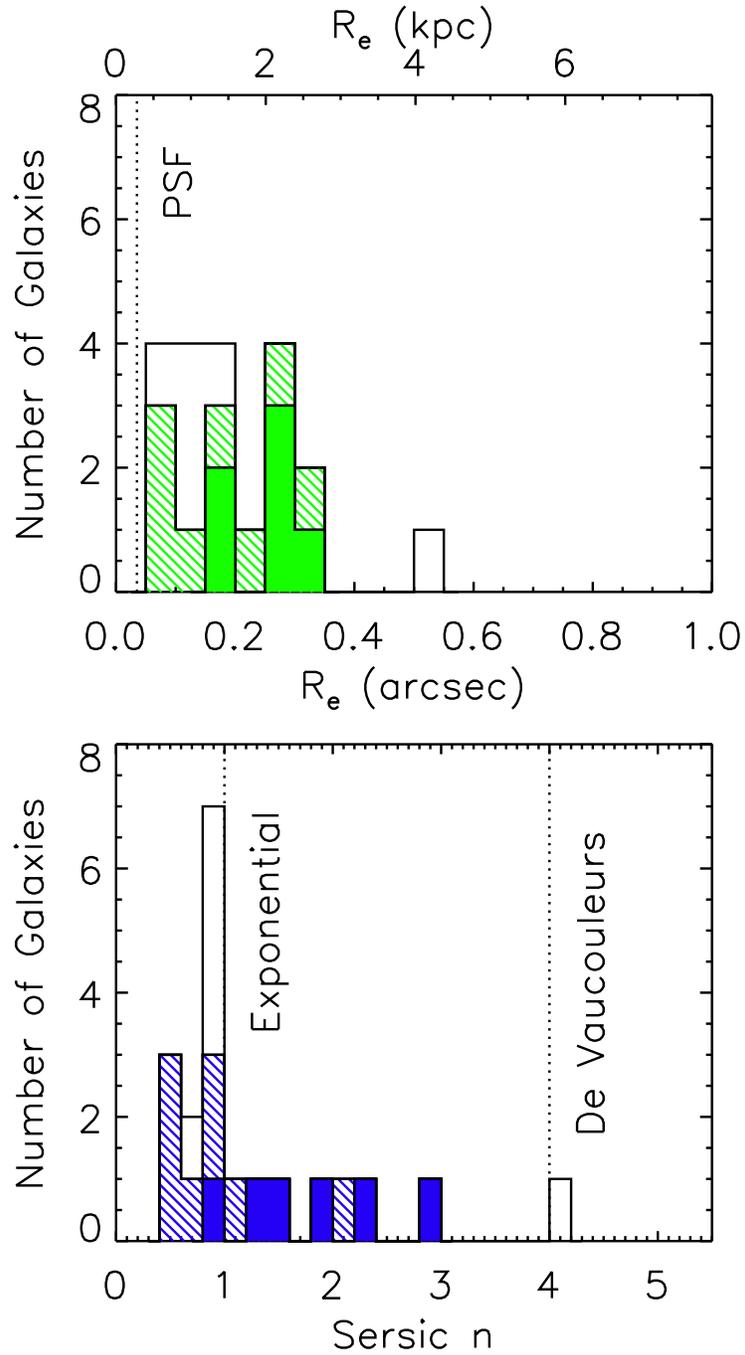}
\caption[GALFIT Sizes and Morphologies of Galaxies Near a \lya\ Nebula.]
{$V_{606}$ morphologies derived using GALFIT for galaxies near LABd05.  
Open histograms contain all sources; solid and hatched histograms 
represent sources in groups ``M1" and ``M2", respectively (Section~\ref{sec:membership}).  
{\bf Top:}  The histogram of galaxy effective radii ($R_{e}$), with the effective 
radius of the ACS PSF shown for reference (dotted line).  
{\bf Bottom:}  The histogram of S\'ersic indices ($n$), 
with the values for exponential ($n=1$) and 
De Vaucouleurs ($n=4$) profiles shown for reference (dotted lines).  
}
\label{fig:sizemorphV}
\end{figure}

These data also reveal clues about the ages and morphologies of the galaxies within LABd05.  
The colors of sources within the system appear to be consistent with young ages ($\approx25-100$~Myr) 
and a range of dust extinctions ($E(B-V)\approx0.0-0.4$).  
The distribution of sizes and S\'ersic indices of the galaxies in the vicinity of LABd05 are shown in Figure~\ref{fig:sizemorphV}, 
with those that are likely associated with LABd05 shown as filled (subset M1) and hatched (subset M2) histograms.  
In general, the galaxies in the vicinity of LABd05 are small ($R_{e}=0.5-3$~kpc) with S\'ersic indices clustering 
around $n\approx1$, indicative of exponential disk morphologies.  

In summary, our analysis suggests that the LABd05 system hosts a large population of young, small, disky, low luminosity galaxies.

\subsubsection{Energy Budget of the Member Galaxies}
\label{sec:energy}

The presence of a large population of compact galaxies within the LABd05 system raises an obvious question: 
are they responsible for powering the \lya\ emission via photoionization?  
We find that the total ionizing luminosity ($L_{200-912\AA}$) of all \numgal\ of the well-detected nearby
galaxies at the system redshift (excluding the obscured AGN, \#\agnid) 
is $8.5\times10^{53}-6.6\times10^{51}$ photons s$^{-1}$ \citep[unreddened, $5-25$~Myr
single bursts][]{bc03}, ignoring the effects of distance, geometric corrections, and non-unity escape fractions.  
This is much less than is required to power the observed \lya\ ($1.7\times10^{55}$ photons s$^{-1}$; Paper~I).  
The fact that all of these galaxies lie tens of kiloparsecs away from the peak of the \lya\ emission 
makes their potential contribution even smaller.  

The only single compact source that could potentially contribute substantially to powering the nebula is the AGN (\#\agnid), 
but simple energetic arguments imply that even it may not be powerful enough to explain all of the \lya\ emission.  
In Paper~I, we found that the total ionizing luminosity of the AGN, estimated from an extrapolation of the mid-infrared SED, 
was $\approx1.8\times10^{54}$ photons s$^{-1}$ and potentially orders of magnitude larger than the contribution from 
the other compact sources.  There is a large associated uncertainty on this estimate due to the difficulty of extrapolating 
the ionizing luminosity from the mid-infrared SED, but if it is accurate, it implies that the AGN can contribute at most 18\% 
of the necessary ionizing photons to explain the \lya\ (Paper~I).  Recently, \citet{colbert11} argued, based on measurements 
of PAH features in IRS spectroscopy, that roughly half of the bolometric luminosity of this mid-infrared source may arise 
from star formation rather than AGN activity.  If so, our estimate of the fraction of ionizing photons contributed by the AGN 
must be reduced by an additional factor of 2.  Thus, for the AGN to be the dominant power source for the \lya\ nebula, 
the geometry of the system must be such that the AGN is highly obscured to our line-of-sight but relatively unobscured 
in the direction of the gas cloud.  A full analysis of the energetics of the system and the dominant power source will be 
addressed in an upcoming paper.

\subsection{An Offset Morphology with No Central Galaxy}
\label{sec:offset}

A key observational result  of this work is that there is no galaxy or AGN at or near the 
center of the \lya\ nebula itself.  All the compact sources visible in the HST imaging 
are offset by $\approx$20~kpc from the peak of the \lya\ emission.  
In fact, the member galaxies actually appear to encircle the nebula (e.g., \#1, 18, 21, \lbgidA, 29, 45, 52, 57), although 
given the small sample size, it is not clear whether or not this effect is significant or merely coincidental.  
Neither the \lya\ nor the diffuse UV continuum components show any significant knots of emission 
that might represent bright star-forming galaxies or luminous stellar clusters.  
The lack of compact components brighter than $3.5\times$10$^{-17}$ \ergscm\ (5$\sigma$)  
in the diffuse $V_{606}$-band light implies that there are no compact sources with 
luminosities brighter than $\sim2\times10^{42}$ \ergss\ ($0.03 L^*$) 
embedded within the central line-emitting region of the nebula.  Similarly, the lack of any obvious substructure in the ACS \lya\ image suggests 
that there are no compact sources forming stars at a rate greater than $\sim1~{\rm M_\odot\ yr^{-1}}$.  
The lack of 24\micron\ emission spatially associated with the \lya\ nebula suggests there is also no 
hidden starburst at the center.  This lack of a central galaxy has important implications for the 
potential power source in LABd05.   

An early model that was proposed to explain the \lya\ nebula phenomenon was shock-heating 
in starburst superwinds.  However, from previous work we know that the \civ\ and \heii\ line ratios 
in LABd05 are not consistent with shock ionization (Paper~I).
More importantly, with no central source present to drive a wind it is extremely difficult to describe 
this system using a simple outflow model.  An outflow driven by one of the other sources would  
need to be severely asymmetric (no secondary \lya\ peak is present on the opposite side) and extend 
more than $\sim$40~kpc in one direction.  

The lack of a central galaxy is similarly problematic for models that attempt to explain large \lya\ nebulae such as LABd05 
as evidence for cold flows.  In this scenario, the \lya\ is powered primarily via gravitational cooling radiation, 
as gas falls into a gravitational potential well along cold filaments, heats collisionally, 
and cools via \lya.  The theoretical expectation, borne out more recently in numerical simulations, is that 
the extended \lya\ emission surrounds a growing galaxy 
\citep{hai00,far01,fur05,yang06,dijk06a,dijk06b,goerdt10,faucher10}.  
This basic prediction of the cold flow model is violated in the case of LABd05.  
In order to reconcile the data with this scenario, we would have to conclude that we are witnessing 
the Ly$\alpha$ nebula at an epoch {\it before} the star clusters within it have dynamically relaxed to 
form a centrally concentrated galaxy.  Even in this case, since the observed centroid of the \lya\ emission in this model 
should be a direct indication of the position of the center of mass of the system, we would need to understand how 
{\it all} of the compact sources --- the luminous, obscured AGN and the \numgal\ other compact galaxies --- 
ended up $\gtrsim20$~kpc away from the center of mass.  

Finally, another hypothesis for explaining the observed properties of \lya\ nebulae is one in which \lya\ photons produced 
by a central source are resonantly scattered in the surrounding gas out to much larger radii.  Recent observations have 
uncovered what appear to be \lya\ scattering halos around continuum-selected LBGs, with \lya\ surface brightness profiles 
similar in shape to those of \lya\ nebulae \citep{stei11}, and have revived the long-standing question of whether \lya\ nebulae 
could simply be scaled up versions of this phenomenon.  
In addition, imaging polarimetry of one of the $z=3.1$ \lya\ nebulae in the SSA22 field (LAB1) resulted in a 
detection of polarization of the \lya\ emission \citep{hayes11}, suggesting that in this case \lya\ photons may be 
scattering from central sources embedded within the nebula \citep[cf.][]{weij10} rather than being produced 
{\it in situ} at large radii.  
While resonant scattering must contribute at some level to the overall extent of \lya\ nebulae, it is not yet 
clear whether this process is the dominant cause of the large sizes observed in all (or even most) \lya\ nebulae.  
Indeed, this explanation is difficult to reconcile with our observations of LABd05. 
While existing imaging polarimetric constraints for LABd05 \citep{pres11} are not sufficient 
to rule out scattering entirely, with no central galaxy or galaxies, there is no obvious source of 
\lya\ photons that could undergo resonant scattering 
to produce the observed \lya\ nebula.  Instead, the \lya\ photons could only be supplied by the AGN or 
compact galaxies at the outskirts.  The fact that the \lya\ is so dramatically offset from these 
potential source(s) of \lya\ photons would require that either the gas distribution or the illumination is severely asymmetric.   

We note that LABd05 is not the only \lya\ nebulae known to lack a central source.  The extensive multi-wavelength 
imaging of the GOODS-S field revealed a \lya-emitting nebula at $z\approx3.16$ with no obvious continuum counterparts.  This fact 
was used by the authors to argue the \lya\ emission is most likely powered by gravitational cooling radiation \citep{nil06}, 
which, if true, would again require the system to be in a very early state prior to the onset of significant star formation 
at or near the center of mass of the system.  
However, the deep HST multiband imaging of the field does reveal a number of compact galaxies within 1-7\arcsec\ of 
the \lya\ emission peak.  While none of the available photometric redshifts for these galaxies closely matches the redshift of 
the \lya\ nebula, four of the brighter cases are consistent with the systemic redshift, given the large redshift error bars, 
and a number of fainter galaxies without reliable photometric redshifts are located within 1-3\arcsec\ of the \lya\ emission peak.  
In the context of what we have learned about LABd05, it is clear that powerful \lya\ nebulae can exist substantially 
offset from all associated continuum sources.  We argue therefore that it remains to be seen whether the \citet{nil06} 
\lya\ nebula is in fact alone.

\subsection{Diffuse Line and Continuum Emission}
\label{sec:diffuse}

The ACS and ground-based broad-band images show clear evidence for a spatially extended diffuse UV continuum component that is 
co-located with the spatially extended \lya\ emission.  In addition, the non-detection in our narrow-band 
$FR601N$ imaging strongly suggests that the \heii\ emission is spatially extended as well ($\minheii\arcsec<$~FWHM~$<1.0\arcsec$).  
What is the origin of these diffuse components?  Given the lack of a central galaxy, there are only two plausible sources: 
in situ spatially distributed star formation and/or the obscured AGN that lies $>20$~kpc away.  
  
One intriguing possibility is that the diffuse UV continuum emission is 
due to star-formation taking place in very small, widely distributed, perhaps dynamically unrelaxed 
regions that are unresolved and unseen by the current HST imaging.  
Possible evidence for this type of extended star formation has been seen in another radio-quiet \lya\ nebula 
\citep[SSA22-LAB1;][]{mat07} as well as in the outskirts of a radio galaxy \citep[MRC 1138-262, 
``the Spiderweb galaxy'';][]{hatch08}.  
This scenario would explain the morphology of the observed diffuse continuum component and 
the presense of ionizing radiation emerging over an extended region.  
The roughly elliptical shape of the UV continuum and \lya\ components, along with the 
velocity profile of the \lya\ reported by Paper I, lead us to speculate that we could 
be observing a large inclined disk exhibiting solid-body rotation.  
Using the central aperture measurements for both components (Table~\ref{tab:galfitdiffuse}), 
we derive similar star formation rates from the \lya\ and diffuse UV continuum emission 
\citep[$8.3\pm0.1$ and $5.6\pm0.5$ M$_{\odot}$ yr$^{-1}$, respectively, within the central 2 arcsec$^{2}$;][]{ken98review}.  
The corresponding rest-frame equivalent width for \lya\ of $\approx200\pm18$\AA, which is plausible for a stellar 
population \citep{char93,mal02,sch03}.  
However, the total diffuse UV continuum flux implies that the expected contribution to 
the ionizing photon flux is $1.7\times10^{53}-1.3\times10^{51}$ photons s$^{-1}$ from a young ($5-25$~Myr), solar 
metallicity, spatially distributed stellar population with a mass of $5.0\times10^{7}-4.6\times10^{8}$ M$_{\odot}$ 
\citep[unreddened single bursts;][]{bc03}.  This is only a fraction of the ionizing flux required to 
power the entire \lya\ nebula ($1.6\times10^{55}$ photons s$^{-1}$; Paper~I), suggesting that a much lower metallicity stellar population 
would be required in this scenario.  The inferred spatial extent of the \heii\ emission could potentially support of this picture; 
however, this would imply a stellar population of extremely young age and low metallicity \citep[$<2$~Myr, $Z<10^{-7} Z_{\odot}$;][]{sch03}.  

The alternative is that the diffuse line and continuum emission are the result of photoionization and scattering from 
the AGN located nearly 20~kpc away from the peak of the diffuse line and continuum components.  Evaluating this scenario 
requires knowledge of the total power of the AGN and the degree of obscuration in the direction of the nebula.  
The bolometric luminosity derived from the infrared SED indicates a powerful AGN \citep[$8.6\times10^{12} L_{\odot}$;][]{yang12}.  
On the other hand, \citet{colbert11} found evidence for strong PAH emission in the mid-infrared spectrum of this source 
and concluded that a significant fraction of the bolometric luminosity likely results from star-formation as well.  
Hence, while an AGN beamed in the direction of the nebula but obscured from our direct view can account for some of the observed diffuse 
light, it is not yet clear what fraction of the observed \lya\ and UV continuum emission can be explained by this scenario. 

Ultimately, understanding the origin of the diffuse continuum components will require better data than are currently available.  
In particular, deep imaging to measure the continuum colors, deep polarization observations to determine what fraction, 
if any, of the continuum light is scattered, and a map of the velocity field measured from a non-resonant line 
(e.g., H$\alpha$ or [OIII]$\lambda$5007) are necessary.  Given the apparent faintness and redshift of this target, 
these observations await JWST.

\subsection{Smooth, Non-filamentary Morphology}
\label{sec:smooth}

Another key morphological result of this work is the finding that the diffuse emission components (\lya\ and UV) in 
LABd05 are remarkably smooth and round.  
In particular, there is no evidence for the kind of bubble-like structures that have been taken as 
evidence for the superwind outflow scenario \citep[e.g.,][]{tani01,mori04,mat04}.  
The diffuse emission is also not particularly clumpy or filamentary, 
in contrast to the predictions of recent cold flow simulations 
that suggest the morphologies of \lya\ nebulae powered by cold accretion should be 
asymmetric and narrow with ``finger-like extensions'' \citep{goerdt10}.  Instead, the \lya\ 
morphology of LABd05 is quite symmetric and well-described by an exponential disk.  While the 
typical axis ratio of the predicted \lya\ nebulae appear \citep[Figures~$7-9$ of][]{goerdt10} 
to be in the range $b/a\approx0.25-0.5$, LABd05 is much less elongated with an axis ratio of 
$0.79$. The \citet{goerdt10} models also suggest that clumps associated with the inflowing 
streams should provide an important contribution to the total luminosity.  Quantitative estimates are 
not given in the paper, but from their Figures~$7-9$ we estimate that there should be $3-4$ 
significant clumps (with surface brightnesses $>0.1$ times that of the central peak) within the 
virial radius ($\approx70$~kpc).  No significant clumps are seen in the \lya\ emission from LABd05.  

\begin{figure}
\center
\includegraphics[angle=0,width=4in]{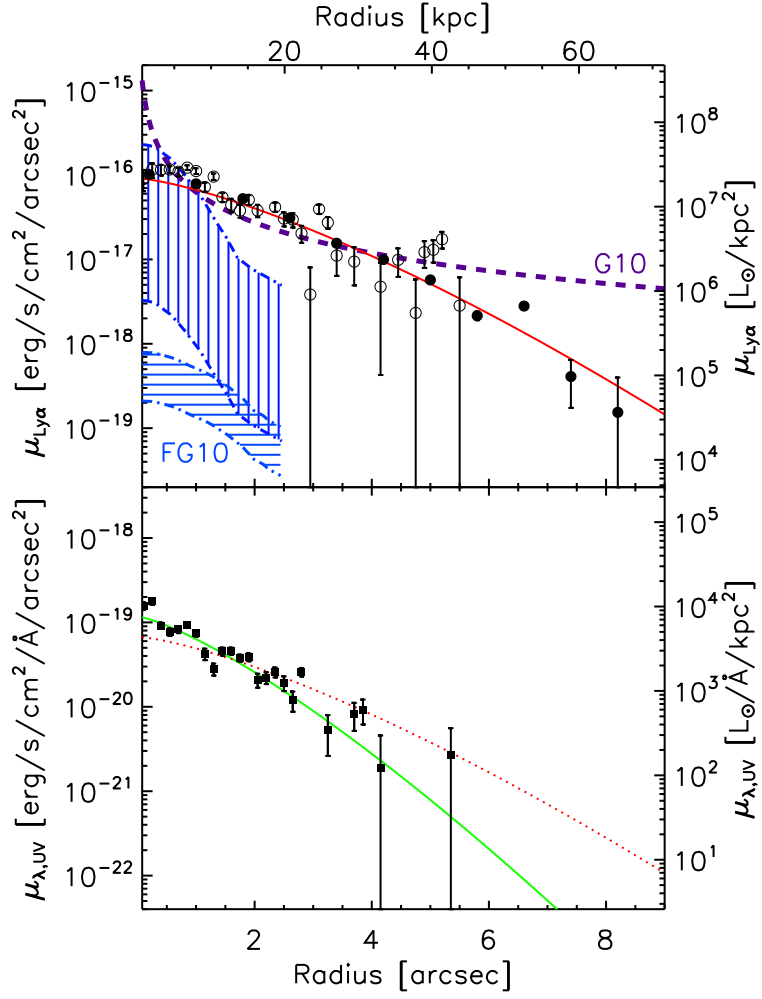}
\caption[Surface Brightness Profiles of LABd05]{
Surface brightness profiles of the \lya\ and diffuse UV continuum emission in LABd05.  
{\bf Top:} \lya\ surface brightness profiles computed using the Subaru \lya\ (filled circles) and ACS \lya\ 
(small open circles) imaging as well as the GALFIT parametric fit to the Subaru \lya\ imaging 
(red line).  Predictions from cold flow simulations are overplotted and labeled with 
`G10' \citep[dashed purple line; the prediction from][scaled to match the total \lya\ luminosity of LABd05 within a 5\arcsec\ radius]{goerdt10} 
and `FG10' \citep[hatched blue regions bounded by dot-dashed lines;][]{faucher10}, 
as discussed in Section~\ref{sec:smooth}.  
The observed profile of LABd05 is not well-described by the predictions from existing cold flow simulations.  
{\bf Bottom:}  $V_{606}$ surface brightness profile of the diffuse UV continuum emission in LABd05 (squares) 
as well as the GALFIT parametric fit (green line).  A GALFIT parametric fit to the Subaru \lya\ imaging from the 
top panel is reproduced for reference (dotted red line), after being scaled to match the total observed flux of 
the diffuse UV continuum within a radius of 3\arcsec.  
The centroids of the \lya\ and diffuse UV continuum emission are offset by $\approx\lyauvoffset$\arcsec.
}
\label{fig:sbprofile1}
\end{figure}

The surface brightness profiles of \lya\ emission from model nebulae are also typically more 
centrally concentrated than that observed for LABd05.  
In Figure~\ref{fig:sbprofile1} we overplot the predicted surface brightness profiles from recent cold flow 
simulations.  The \citet{goerdt10} prediction of a $r^{-1.2}$ power law, which we have scaled to 
match the total observed \lya\ flux for LABd05 inside a radius of 5\arcsec, is a poor fit 
to the shape of the observed surface brightness profile.  The \citet{faucher10} predictions shown 
are based on their two most realistic treatments for \lya\ emission from gravitational 
cooling (their models \#7 and \#9, which include prescriptions for self-shielding).  The lower 
bound of each region shown corresponds to the prediction for a fiducial $2.5\times10^{11}$ M$_{\odot}$ 
halo mass model at $z=3$; the upper bound of each region is the same profile scaled up to a halo mass of 
$10^{13}$ M$_{\odot}$ based on the predicted $L_{Ly\alpha}$-M$_{halo}$ relation \citep{faucher10} and 
under the na\"ive assumption that the profile shape is constant as a function of halo mass.  
Model \#7 can in principle reach the peak \lya\ surface brightness we observe in LABd05, assuming a 
sufficiently massive halo, but the profile shape is much more centrally concentrated than is observed.  
Model \#9 is orders of magnitude too faint, even for the most massive halos. 
 
We note that although the expected luminosity scaling is the most basic output from models of \lya\ nebulae 
powered by cold accretion, this has turned out to be particularly difficult to predict robustly.  
Early models suggested that the \lya\ emission from gravitational cooling should be similar to 
what is observed in \lya\ nebulae \citep[e.g.,][]{yang06,goerdt10}.  However, more recent work has argued 
that these \lya\ nebula luminosity predictions may be orders of magnitude too high due to the effects 
of self-shielding and that cooling radiation alone is an unlikely explanation for the \lya\ emission of 
the most luminous \lya\ nebulae \citep{faucher10}.  The question of the predicted \lya\ luminosity from 
cold accretion is still a matter of some debate, but interestingly our analysis has shown that even if the 
question of the luminosity scaling is ignored, key morphological discrepancies remain between existing 
cold flow models and what is seen in LABd05.

\subsection{Similar UV and Ly$\alpha$ Surface Brightness Profiles}
\label{sec:sbprofile}

Due to the effects of resonant scattering, there is a generic expectation that high redshift sources of 
\lya\ emission should be surrounded by low surface brightness 
halos of resonantly scattered \lya\ emission \citep[e.g.,][]{loeb99,zheng10}.  After a number of observational studies 
uncovered possible hints of this extended emission in samples of \lya-emitting galaxies \citep[e.g.,][]{hay04,ono10}, 
\citet{stei11} used a stacking analysis to demonstrate convincingly that extended \lya\ halos appear to exist 
around all classes of star-forming galaxies.  They noted the large extent of these \lya\ halos relative to the much 
more compact UV cores and pointed out the similarities in surface brightness profile shape between the \lya\ halos 
around stacked continuum-selected LBGs and stacked \lya\ nebulae in the same field.  They concluded that, 
if one could image deeply enough ($F_{Ly\alpha}\approx10^{-19}$ \ergscm), all continuum-selected LBGs would be 
classified as extended \lya\ nebulae.  The obvious question then becomes: are giant \lya\ nebulae simply scaled 
up versions of this phenomenon with the large extent driven simply by resonant scattering 
of \lya\ photons from a single or several central sources?   

\begin{figure}
\center
\includegraphics[angle=0,width=4in]{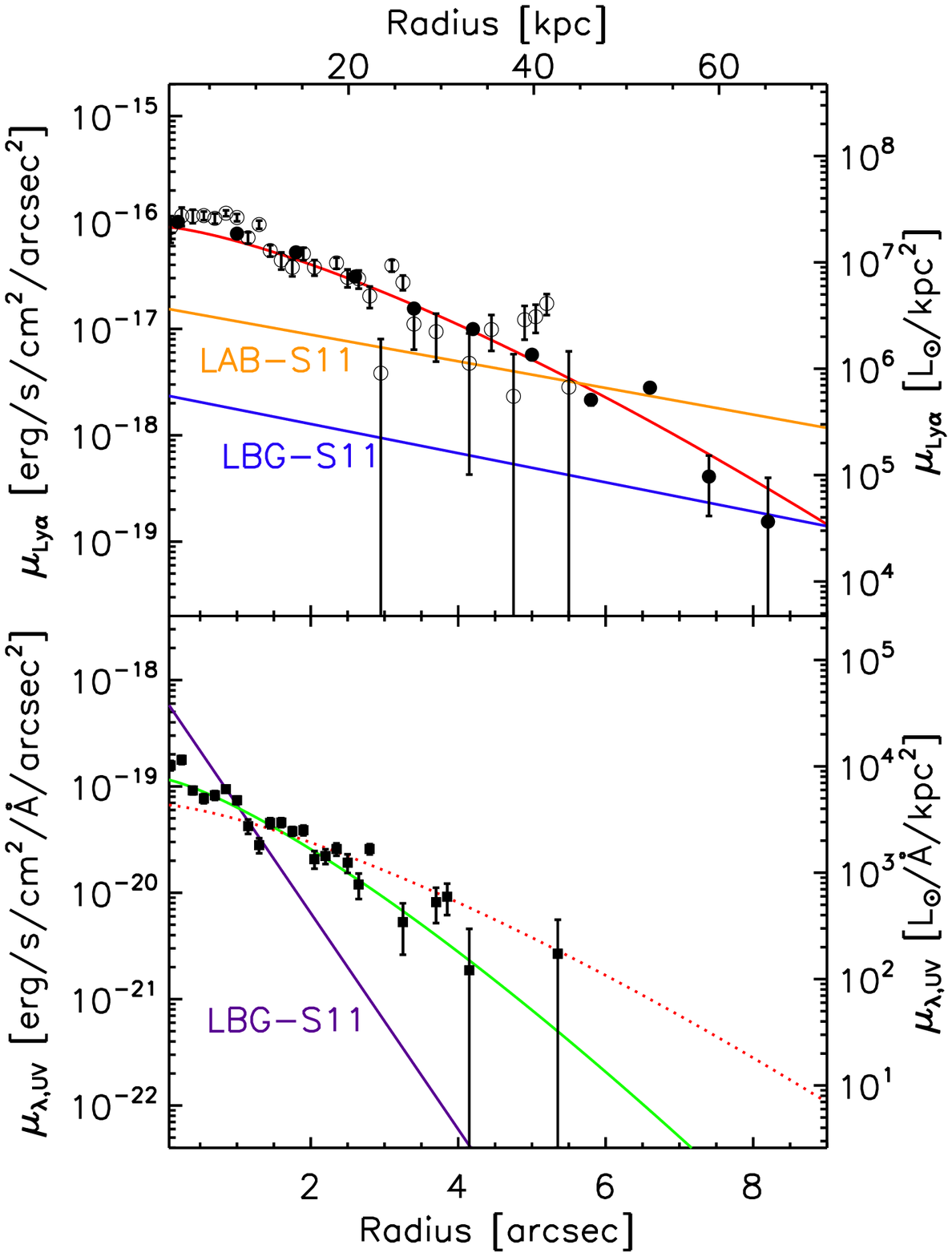}
\caption[Surface Brightness Profiles of LABd05]{
Surface brightness profiles of the \lya\ and diffuse UV continuum emission in LABd05, as shown in 
Figure~\ref{fig:sbprofile1}.  
{\bf Top:}  The stacked \lya\ surface brightness profiles of continuum-selected galaxies (`LBG-S11'; blue line) and of 
\lya\ nebulae (`LAB-S11'; orange line) are overplotted.  
{\bf Bottom:} The stacked continuum surface brightness profile of continuum-selected galaxies 
from \citet{stei11} is overplotted (`LBG-S11'; purple line).  
The stacked LBGs show compact UV emission surrounded by an extended \lya\ halo, likely evidence for 
resonant scattering of \lya\ photons generated in the core; in contrast the \lya\ and UV surface 
brightness profiles for LABd05, while not identical, are strikingly similar in shape and extent. 
}
\label{fig:sbprofile2}
\end{figure}

In the case of LABd05, it does not appear that the large \lya\ extent can be explained simply as a result of resonant scattering.  
The lack of an obvious central source of \lya\ photons is the first challenge 
(as discussed in Section~\ref{sec:offset}), but another inconsistency appears when considering the \lya\ and UV 
surface brightness profiles (Figure~\ref{fig:sbprofile2}).  Instead of a compact UV core surrounded by an extended \lya\ 
halo, the UV emission in LABd05 is nearly as spatially extended as the \lya, a clear indication that the the nebula's large 
size is primarily the result of some other mechanism or geometry.  
Furthermore, the observed \lya\ and diffuse UV continuum 
surface brightness profiles are remarkably similar in their properties.  
They both show similar radial distributions, are well-described by nearly-exponential disks, 
and have approximately elliptical shapes with similar 
axial ratios (Table~\ref{tab:galfitdiffuse}).  
In comparison, the stacked UV and \lya\ profiles derived for continuum-selected 
star-forming galaxies differ significantly \citep[Figure~\ref{fig:sbprofile2};][]{stei11}.  
Thus, while it seems likely that 
resonant scattering can explain some of the remaining differences between the two components, it does not appear that the large 
spatial extent of LABd05 is solely a result of \lya\ resonant scattering.

\subsection{A Galaxy Group in Formation?}
\label{sec:implications}

LABd05 is unique only because of the existence of deep and high-spatial resolution broad- and narrow-band imaging data. 
These data provide many pieces of evidence in support of the idea that this is a young, forming system.  
There are numerous small, low-luminosity, disky galaxies, many of which have very blue colors. 
Even the reddest objects have colors that are consistent with ages of less than a few hundred million years, 
perhaps only a hundred million years if they are modestly reddened (i.e., $E(B-V)\approx 0.4$).  
The large and luminous \lya\ halo and the 
detection of faint, diffuse UV continuum emission in the region also suggest that the system is energetically young.  
In addition, the fact that \nummone\ of the compact galaxies that are likely to be associated with the system 
(i.e., roughly half the candidate members, with a total luminosity of 1.2~$L^{*}$) lie within a small projected 
area $\approx$30~kpc in diameter suggests that the system may also be {\it dynamically young}, the dynamical time 
for this region being only $\sim14$~Myr.\footnote{This calculation assumes that $L^{*}$ corresponds to a mass of 
$\approx2\times10^{10}$ M$^{\odot}$ \citep{shap05,red06,erb06,red09} and a stellar-to-halo mass ratio at this 
mass of $10^{-2}$ for $z=3$ \citep{most10}.} 

We speculate that this giant \lya\ nebula is the progenitor of a galaxy group, witnessed in the process of formation.  
Under this assumption, the luminosity distribution of the member galaxies can provide a unique perspective on the 
``initial luminosity function'' of galaxies.  Spectroscopic redshift measurements of the galaxies within the system 
will be key to confirming membership and determining the dynamical mass of the system.  While we only have spectroscopic redshifts 
for 3 sources in the region, we have argued both from the excess of galaxies in the vicinity of the nebula and from 
the colors of these galaxies that most of the compact objects observed are likely members of the system.  By assuming that all 
the galaxies that lie above the dashed line in Figure~\ref{fig:VJHcolor2} are members, we have constructed a luminosity 
function for the system, as shown in Figure~\ref{fig:lumfunc} (bottom panel).  
Summing all the UV luminosity contributed by the candidate member galaxies and the diffuse continuum results in a total of 
$\approx$23.4 AB mag in the $V_{606}$ band, or $\approx 3L^{*}$.  It is therefore possible that this system could evolve 
into a small group, with the smaller galaxies merging into larger systems over a few dynamical times.  

Clearly better data are needed, both to confirm the system membership and to measure the stellar masses of the member galaxies 
more robustly.  Nevertheless, LABd05 provides the tantalizing hope that detailed studies of more such systems, even 
statistical studies, can result in a determination of the initial mass function of galaxies, analogous to the manner in 
which studies of stellar clusters in our own Galaxy have yielded the stellar initial mass function.
While LABd05 is only one source, we note that there are other \lya\ nebulae that 
appear to be similar in morphology.  For example, one of the \lya\ nebulae in the SSA22 
field at $z\approx3.1$ (LAB1) shows multiple embedded galaxies and a hint of diffuse UV continuum 
in between the galaxies in ground-based data \citep{mat07}.  
As discussed in Section~\ref{sec:offset}, the \lya\ nebula found in the GOODS-S field at $z\approx3.16$ 
has no central continuum counterpart \citep{nil06}.  Deep HST imaging of a 
larger sample of giant \lya\ nebulae will be important for understanding the 
extent to which the morphology and galaxy properties observed in LABd05 are 
characteristic of \lya\ nebula systems in general.

\section{Conclusions}
\label{sec:conclusions}

Using high resolution HST imaging, we have taken a census of all the compact sources 
within a large \lya\ nebula at $z\approx2.656$.  We find that the \lya\ nebula system 
contains numerous compact, young, disky galaxies and an obscured AGN 
that are all located tens of kiloparsecs from the peak of the 
\lya\ emission and provide a negligible contribution to the ionization of the nebula.  
The observed luminosity function shows that the compact sources within the system are predominantly 
low luminosity ($\sim0.1 L^*$) galaxies, highly suggestive of a galaxy forming environment.  
The large-scale morphology of the system is characterized by the lack of a central galaxy 
at or near the peak of the \lya\ nebula and the presence of several nearly coincident, 
smooth, and spatially extended emission components (\lya, \heii, and diffuse UV continuum).  
These morphological results --- in particular the lack of a central galaxy and the offset 
morphology --- disfavor models of outflows, cold flows, and resonant scattering halos, suggesting 
that while these phenomena may be present, they are not sufficient to explain the powering 
and the large extent of giant \lya\ nebulae.  
Based on these observations, we speculate that large \lya\ nebulae are progenitors of low-redshift 
galaxy groups or low-mass clusters.

\acknowledgments

The authors would like to thank Kate Brand, Galina Soutchkova, and Sangeeta Malhotra for their assistance 
with the HST observation planning and execution.  We are grateful to Crystal Martin, Kristian Finlator, 
Avi Loeb, Dan Weedman, Tommaso Treu, Matt Auger, and the anonymous referee for useful discussions and 
suggestions.  M.~K.~M.~P. acknowledges support from an NSF Graduate Research Fellowship, 
a P.E.O. Fellowship, and a TABASGO Prize Postdoctoral Fellowship.  
This work was based on observations made with the NASA/ESA Hubble Space Telescope 
(HST Cycle 14; GO\#10591), obtained at the Space Telescope Science Institute, which 
is operated by the Association of Universities for Research in Astronomy, Inc., 
under NASA contract NAS 5-26555.  
This work is also based in part on data collected at the Subaru Telescope, which is 
operated by the National Astronomical Observatory of Japan, and on data from the 
NOAO Deep Wide-Field Survey (B. Jannuzi, A. Dey) as distributed by the NOAO Science Archive.  
A.~D. and B.~T.~J.'s research activities are supported by NOAO.  NOAO is operated by the Association of Universities 
for Research in Astronomy (AURA), Inc. under a cooperative agreement with the National Science Foundation. 
V.~D. and B.~T.~S. are supported by the Spitzer Space Telescope project, which is managed 
by JPL on behalf of NASA.

\facility{{\it Facilities:} HST, Subaru, Mayall}


\begin{deluxetable}{cccccc}
\tabletypesize{\scriptsize}
\tablecaption{HST Observations of LABd05} 
\tablewidth{0pt}
\tablehead{
\colhead{Instrument} & \colhead{Filter} & \colhead{Exposure Time} & \colhead{$\lambda_{C}$} & \colhead{Bandpass Width} & \colhead{Restframe $\lambda$} \\ 
 &  & (min) &  &  &  at $z\approx2.656$ } 
\startdata
HST/ACS & FR462N ([O~\textsc{ii}] outer ramp) &  216 & 4448~\AA\tablenotemark{a}  & 89~\AA\ & Ly$\alpha\lambda$1216 \\ 
HST/ACS & F606W (Broad-band $V$) & 129  & 5907~\AA\  & 2342~\AA\  & 1295-1936~\AA\ \\ 
HST/ACS & FR601N ([O~\textsc{iii}] outer ramp) & 129  & 5998~\AA\tablenotemark{a} & 120~\AA\ & He~\textsc{ii}$\lambda$1640 \\ 
HST/NICMOS NIC2 & F110W (Broad-band $J$) & 281  & 1.1~$\mu$m  & 0.6~$\mu$m & 2188-3829~\AA\ \\ 
HST/NICMOS NIC2 & F160W (Broad-band $H$) & 281 & 1.6~$\mu$m  & 0.4~$\mu$m &  3829-4923~\AA\ \\ 
\enddata
\tablenotetext{a}{Ramp filters $FR462N$ and $FR601N$ were centered on \lya\ and \heii$\lambda$1640, respectively, at \zblob\ during these observations.}
\label{tab:hstobs}
\end{deluxetable}

\begin{deluxetable}{cccc}
\tabletypesize{\scriptsize}
\tablecaption{Astrometric Uncertainty\tablenotemark{a}}
\tablewidth{0pt}
\tablehead{
\colhead{Band} & \colhead{$N_{obj}$\tablenotemark{b}} & \colhead{$\sigma_{\alpha}$} & \colhead{$\sigma_{\delta}$} \\
 &  & (arcsec)  &  (arcsec) }
\startdata
 NDWFS $B_{W}$ & - & - & - \\
           Subaru $IA445$  &  5062  & 0.10  & 0.09  \\
           ACS Ly$\alpha$  &     8  & 0.17  & 0.13  \\
      ACS He$\textsc{ii}$  &    18  & 0.12  & 0.10  \\
            ACS $V_{606}$  &   402  & 0.12  & 0.09  \\
         NICMOS $J_{110}$  &     6  & 0.12  & 0.09  \\
         NICMOS $H_{160}$  &     7  & 0.12  & 0.09  \\
\enddata
\tablenotetext{a}{Astrometric uncertainty relative to the NDWFS \bw\ image.}
\tablenotetext{b}{Number of common sources used to compute astrometric correction.}
\label{tab:hstreg}
\end{deluxetable}

\begin{deluxetable}{ccccccccc}
\tabletypesize{\scriptsize}
\tablecaption{Properties of Compact Sources within LABd05}
\tablewidth{0pt}
\tablehead{
\colhead{ID} & \colhead{Right Ascension} & \colhead{Declination} & \colhead{R$_{e}$\tablenotemark{a}} & \colhead{$n$\tablenotemark{a}} & \colhead{$m_{V}$\tablenotemark{b}} & \colhead{$V_{606}-J_{110}$\tablenotemark{b}} & \colhead{$J_{110}-H_{160}$\tablenotemark{b}} & \colhead{System\tablenotemark{c}} \\
 & (hours) & (degrees) & (arcsec) &  & (AB) & (AB) & (AB) & Member?}
\startdata
   36 &    14:34:10.981 &     33:17:32.48 &   0.39 &   19.9 &  28.24$\pm$ 0.38 &   1.62$\pm$ 0.43 &   1.69$\pm$ 0.22 &                            M1 (AGN)        \\
   26 &    14:34:11.036 &     33:17:34.47 &   0.17 &    1.0 &  25.36$\pm$ 0.03 &   0.09$\pm$ 0.07 &   0.32$\pm$ 0.10 &                            M1 (LBG)        \\
   59 &    14:34:11.041 &     33:17:34.14 &   0.27 &    1.4 &  26.15$\pm$ 0.06 &   0.03$\pm$ 0.15 &   0.36$\pm$ 0.20 &                            M1 (LBG)        \\
   28 &    14:34:10.913 &     33:17:33.80 &   0.15 &    2.0 &  26.86$\pm$ 0.11 &  -0.34$\pm$ 0.37 &   1.00$\pm$ 0.42 &                                  M1        \\
   32 &    14:34:10.990 &     33:17:32.89 &   0.31 &    1.4 &  27.17$\pm$ 0.15 &   0.20$\pm$ 0.32 &   1.10$\pm$ 0.33 &                                  M1        \\
   33 &    14:34:11.025 &     33:17:32.70 & - & - &  27.61$\pm$ 0.21 &   0.95$\pm$ 0.30 &   1.10$\pm$ 0.25 &                                  M1        \\
   34 &    14:34:11.085 &     33:17:33.02 & - & - &  27.98$\pm$ 0.30 &   1.35$\pm$ 0.37 &   1.30$\pm$ 0.24 &                                  M1        \\
   35 &    14:34:11.059 &     33:17:32.76 &   0.29 &    2.8 &  26.46$\pm$ 0.08 &   0.72$\pm$ 0.12 &   0.94$\pm$ 0.12 &                                  M1        \\
   37 &    14:34:11.002 &     33:17:32.35 &   0.27 &    2.2 &  26.80$\pm$ 0.11 &   1.10$\pm$ 0.14 &   1.20$\pm$ 0.11 &                                  M1        \\
    1 &    14:34:11.036 &     33:17:25.76 &   0.31 &    2.0 &  26.63$\pm$ 0.09 &   0.96$\pm$ 0.13 &   0.79$\pm$ 0.11 &                                  M2        \\
   18 &    14:34:10.764 &     33:17:36.75 &   0.15 &    0.4 &  27.29$\pm$ 0.16 &   0.67$\pm$ 0.26 &   0.51$\pm$ 0.30 &                                  M2        \\
   21 &    14:34:10.956 &     33:17:36.06 &   0.26 &    0.9 &  26.73$\pm$ 0.10 &   0.68$\pm$ 0.16 &   0.48$\pm$ 0.18 &                                  M2        \\
   24 &    14:34:10.926 &     33:17:34.81 &   0.23 &    0.8 &  26.46$\pm$ 0.08 &   0.34$\pm$ 0.15 &   0.32$\pm$ 0.21 &                                  M2        \\
   29 &    14:34:10.743 &     33:17:33.85 &   0.09 &    1.1 &  26.81$\pm$ 0.11 &   0.31$\pm$ 0.21 &   0.63$\pm$ 0.25 &                                  M2        \\
   45 &    14:34:11.258 &     33:17:30.24 &   0.10 &    0.9 &  27.59$\pm$ 0.21 &   1.71$\pm$ 0.24 &   1.77$\pm$ 0.11 &                                  M2        \\
   52 &    14:34:10.845 &     33:17:27.10 &   0.11 &    0.6 &  26.69$\pm$ 0.10 &   0.24$\pm$ 0.20 &   0.30$\pm$ 0.29 &                                  M2        \\
   57 &    14:34:11.039 &     33:17:25.56 &   0.08 &    0.5 &  26.61$\pm$ 0.09 &   0.94$\pm$ 0.13 &   0.87$\pm$ 0.11 &                                  M2        \\
    2 &    14:34:10.997 &     33:17:25.98 &   0.11 &    0.4 &  28.04$\pm$  0.32 & - & - &                                   -        \\
   27 &    14:34:10.897 &     33:17:34.12 &   0.21 &    0.2 &  27.97$\pm$  0.30 & - & - &                                   -        \\
   38 &    14:34:10.902 &     33:17:32.29 &   0.20 &    2.6 &  27.62$\pm$  0.22 & - & - &                                   -        \\
   43 &    14:34:11.015 &     33:17:31.17 &   0.16 &    0.2 &  27.54$\pm$  0.20 & - & - &                                   -        \\
   47 &    14:34:10.544 &     33:17:29.10 &   0.17 &    0.7 &  27.77$\pm$  0.25 & - & - &                                   -        \\
   48 &    14:34:10.551 &     33:17:28.61 &   0.16 &    0.8 &  27.76$\pm$  0.25 & - & - &                                   -        \\
   49 &    14:34:11.412 &     33:17:29.15 &   0.08 &    0.7 &  27.47$\pm$  0.19 & - & - &                                   -        \\
    9 &    14:34:10.955 &     33:17:38.49 &   0.17 &    0.9 &  27.19$\pm$ 0.15 &   0.82$\pm$ 0.22 &   0.01$\pm$ 0.32 &                                  NM        \\
   13 &    14:34:11.251 &     33:17:37.90 &   0.09 &    0.9 &  27.18$\pm$ 0.15 &   0.67$\pm$ 0.24 &  -0.29$\pm$ 0.45 &                                  NM        \\
   40 &    14:34:10.555 &     33:17:31.92 &   0.11 &    0.7 &  27.46$\pm$ 0.19 &   0.79$\pm$ 0.29 &  -0.42$\pm$ 0.57 &                                  NM        \\
   58 &    14:34:10.936 &     33:17:36.24 &   0.50 &    4.1 &  27.53$\pm$ 0.20 &   1.13$\pm$ 0.26 &   0.05$\pm$ 0.32 &                                  NM        \\
   46 &    14:34:10.850 &     33:17:29.85 &   0.11 &    0.9 &  25.49$\pm$ 0.04 &  -0.10$\pm$ 0.09 &   0.30$\pm$ 0.13 &       NM ($z=3.2$)\tablenotemark{d}        \\
   60 &    14:34:10.854 &     33:17:30.05 &   0.13 &    1.0 &  25.97$\pm$ 0.05 &   0.05$\pm$ 0.12 &   0.24$\pm$ 0.18 &       NM ($z=3.2$)\tablenotemark{d}        \\
\enddata
\tablenotetext{a}{Morphological measurements $R_{e}$ and $n$ denote the effective radius and S\'ersic index as measured by GALFIT (Section~\ref{sec:morph}).}
\tablenotetext{b}{Aperture magnitudes for the $V_{606}$, $J_{110}$, and $H_{160}$ bands were computed using \apsize\arcsec\ diameter apertures and aperture corrections of \apcorr.}
\tablenotetext{c}{Membership categories based on optical/NIR colors and proximity (see Section~\ref{sec:membership}): 
``M1" - likely system member inside a radius of \lbgdist\arcsec\ from the AGN (\#\agnid); ``M2" - likely system member but beyond a radius of \lbgdist\arcsec\ from the AGN (\#\agnid); ``NM" - likely non-member.  Objects with membership confirmed by a spectroscopic redshift are indicated with parenthetical remarks.  Sources with no membership designation were not well-detected in one or more bands.}
\tablenotetext{d}{$V_{606}$ aperture magnitude uncorrected for contamination by \lya\ emission at $z\approx3.2$ ($F_{Ly\alpha}=5.15\times10^{-17}$ \ergscm; Paper~I).}
\label{tab:galfitcompact}
\end{deluxetable}

\begin{deluxetable}{cccccc}
\tabletypesize{\scriptsize}
\tablecaption{Properties of Diffuse Components within LABd05}
\tablewidth{0pt}
\tablehead{
&  & \colhead{Diffuse Continuum} & & \colhead{Diffuse \lya} & \colhead{Diffuse \heii} \\
Band & $V_{606}$ & $J_{110}$ & $H_{160}$ & $IA445$ &  \\
Restframe wavelength (\AA) & 1298-1939\AA\ & 2298-3720\AA\ & 3829-4923\AA\ & 1640-1716\AA\ & } 
\startdata
GALFIT Parameters\tablenotemark{a}& & & & &  \\
\hline
 & & & & & \\
Right Ascension (hours) & 14:34:10.940 & & & 14:34:10.986 &   14:34:10.925\tablenotemark{b}   \\
Declination (degrees) & +33:17:29.87 & & & +33:17:30.43 & +33:17:29.92\tablenotemark{b} \\
       Size (arcsec) & $R_{e}=  1.67\pm  0.08     $ & & & $R_{e}=  2.48\pm  0.03 $ & FWHM~$=\minheii-\maxheii$\tablenotemark{c} \\
      S\'ersic Index ($n$) & $  0.77\pm  0.05           $ & & & $  0.73\pm  0.01    $ & \\
          Axis Ratio (b/a) & $  0.77\pm  0.03           $ & & & $  0.79\pm  0.01    $ & \\
  Position Angle (\degree) & $-43.49\pm  5.48           $ & & & $-15.51\pm  1.24    $ & \\
      Total Magnitude (AB) & $ 23.81\pm  0.04           $ & & & $ 21.99\pm  0.01    $ & \\
Total Luminosity & & & &  &  \\
          ($10^{28}$ \ergss $Hz^{-1}$) & $L_{\nu}= 17.13\pm  0.63                   $ &  &  &  &  \\
          ($10^{42}$ \ergss) &  &  &  & $L_{\lya}=101.61\pm  0.94             $ &  \\
 & & & & & \\
\hline
 & & & & & \\
\colhead{Aperture Photometry\tablenotemark{d}} & & & & &  \\
\hline
 & & & & & \\
  Magnitude (AB) & $ 25.39\pm  0.09    $ & $> 25.75                        $ & $> 25.79                         $ & $ 24.60\pm  0.01   $ &  \\
Luminosity & & & & &  \\
          ($10^{28}$ \ergss $Hz^{-1}$) & $L_{\nu}=  4.00\pm  0.35                $ &  $L_{\nu}<  2.87                 $ & $L_{\nu}<  2.77                                                                $  &  &  \\
            ($10^{42}$ \ergss) & & &  & $L_{\lya}=  9.15\pm  0.11    $ &  $L_{\heii}=  2.35\pm  0.02               $\tablenotemark{e}\\
\enddata
\tablenotetext{a}{Errors on morphological parameters are formal fitting errors reported by GALFIT but do not include errors resulting from the continuum subtraction or previous fitting of embedded compact sources.  They therefore likely underestimate the true uncertainty in fitting this complex system.}
\tablenotetext{b}{Approximate position of the \heii\ source measured from ground-based spectroscopic data (Paper~I).}
\tablenotetext{c}{Limits on the size derived from ACS \heii\ imaging and ground-based spectroscopic data, as described in Section~\ref{sec:heii}.}
\tablenotetext{d}{Aperture photometry measured at the center of the nebula within a small \diffapsize\arcsec\ diameter aperture, chosen to avoid nearby compact sources.  No aperture corrections were applied.  Limits are 3$\sigma$ values.}
\tablenotetext{e}{\heii\ luminosity measured within a $4.5\times1.5$\arcsec\ spectroscopic slit assuming $z=2.6562$ (Paper~I).}
\label{tab:galfitdiffuse}
\end{deluxetable}

\clearpage

\clearpage

\appendix
\section{Postage Stamps}
\label{appendixA}

\begin{figure}
\center
\includegraphics[angle=0,width=6.5in]{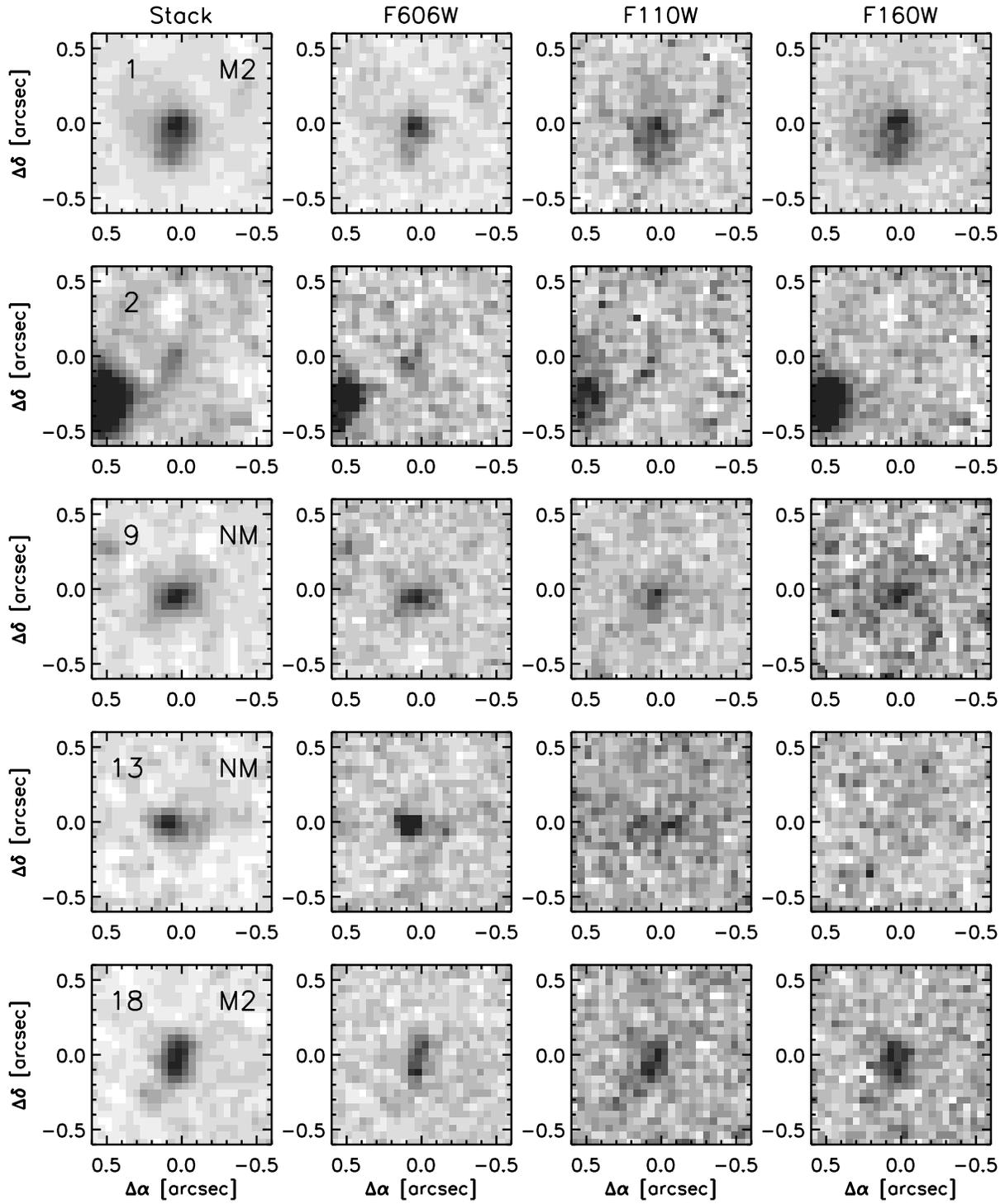}
\caption[Postage Stamps]
{Postage stamp images of the compact sources in the vicinity of LABd05.
}
\label{fig:imappendix1}
\end{figure}

\begin{figure}
\center
\includegraphics[angle=0,width=6.5in]{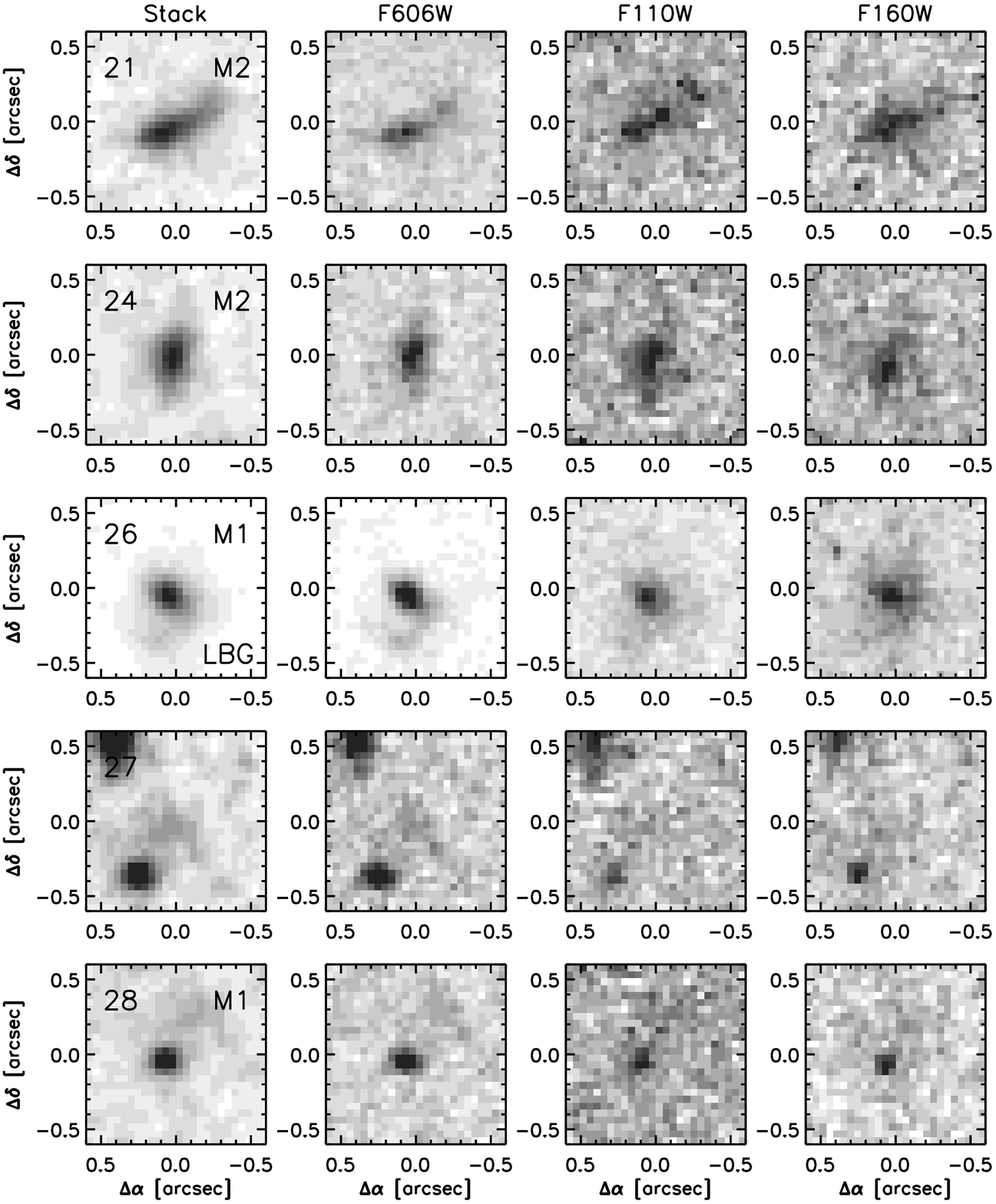}
\caption[Postage Stamps]
{Postage stamp images of the compact sources in the vicinity of LABd05.
}
\label{fig:imappendix2}
\end{figure}

\begin{figure}
\center
\includegraphics[angle=0,width=6.5in]{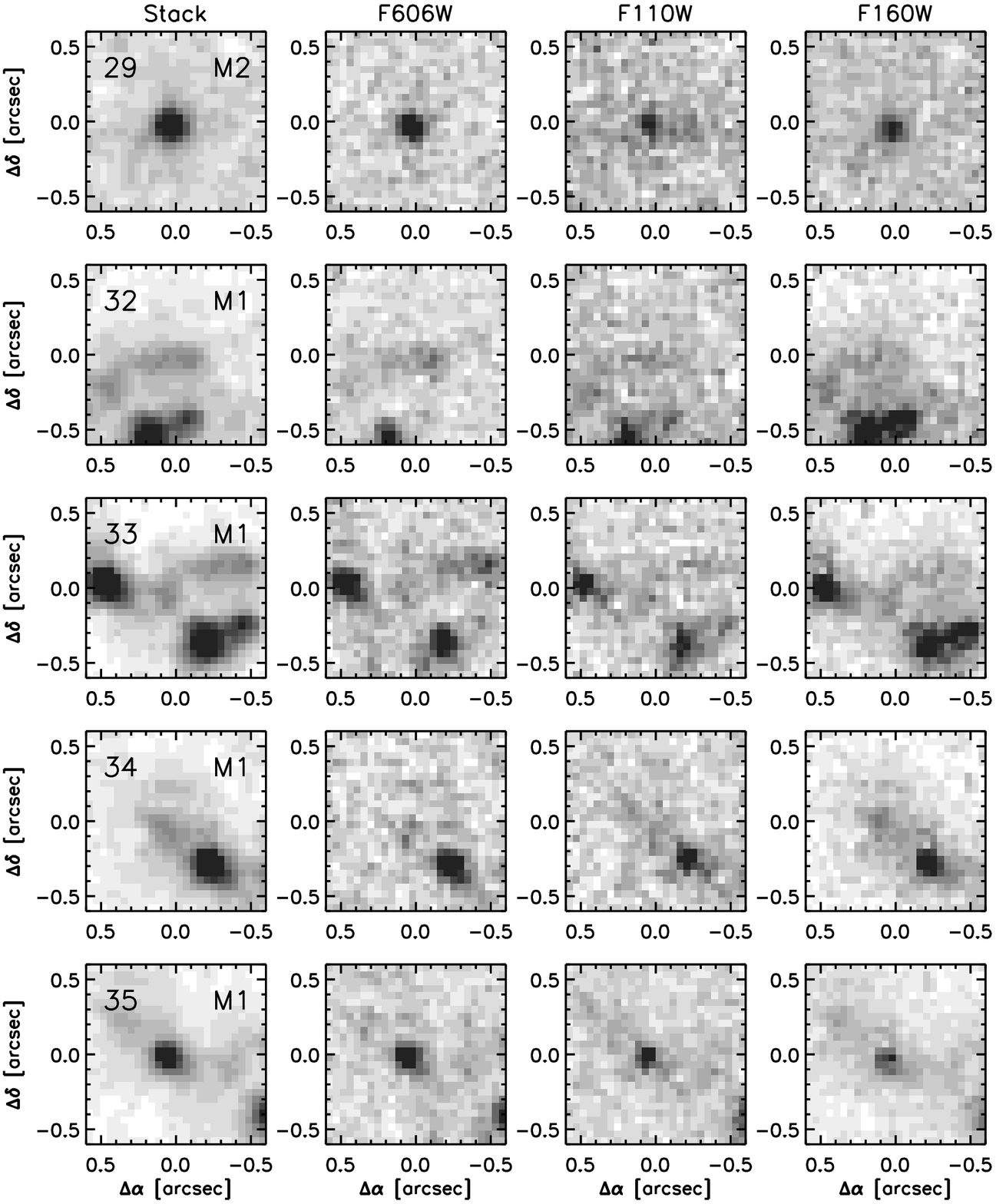}
\caption[Postage Stamps]
{Postage stamp images of the compact sources in the vicinity of LABd05.
}
\label{fig:imappendix3}
\end{figure}

\begin{figure}
\center
\includegraphics[angle=0,width=6.5in]{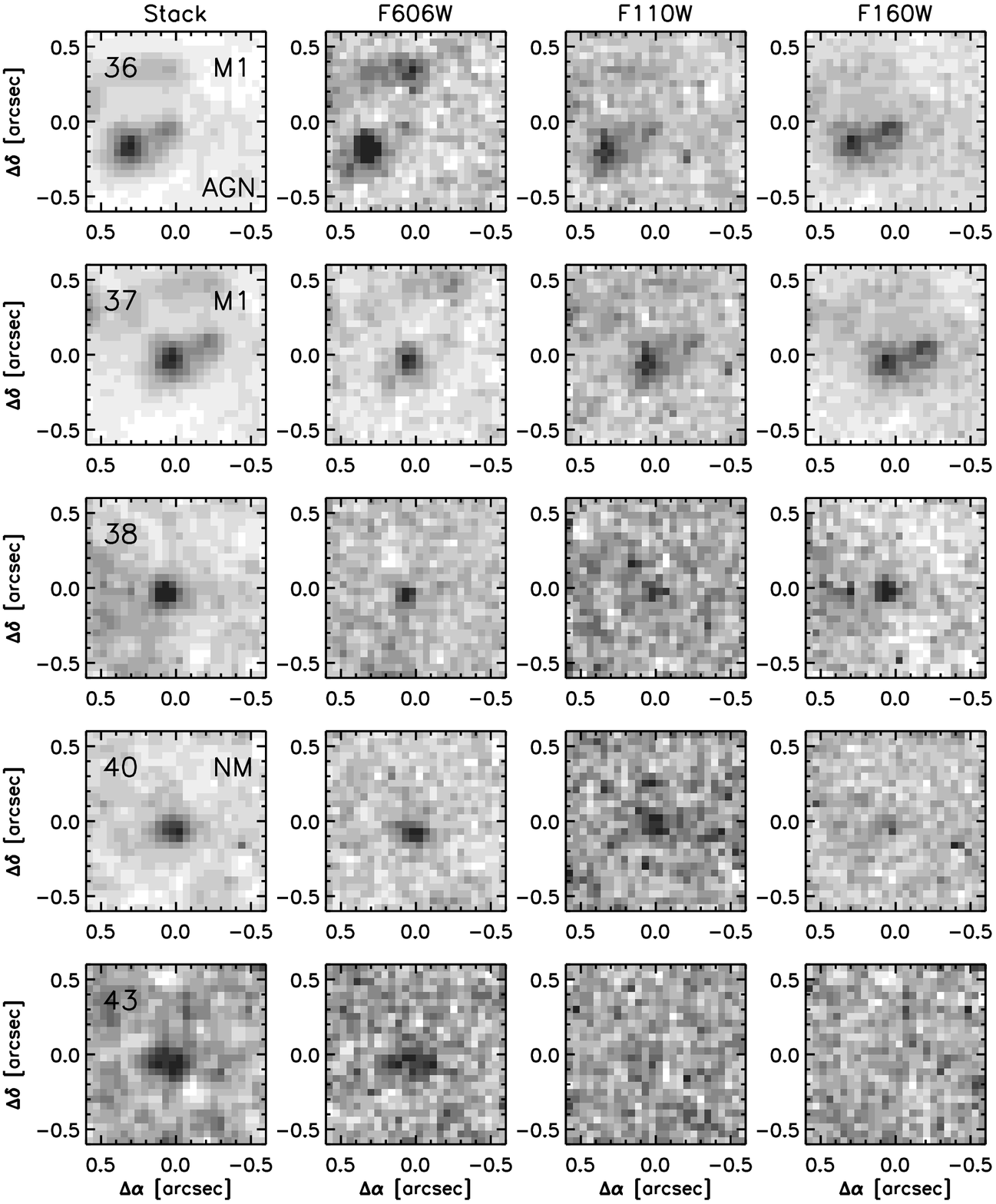}
\caption[Postage Stamps]
{Postage stamp images of the compact sources in the vicinity of LABd05.
}
\label{fig:imappendix4}
\end{figure}

\begin{figure}
\center
\includegraphics[angle=0,width=6.5in]{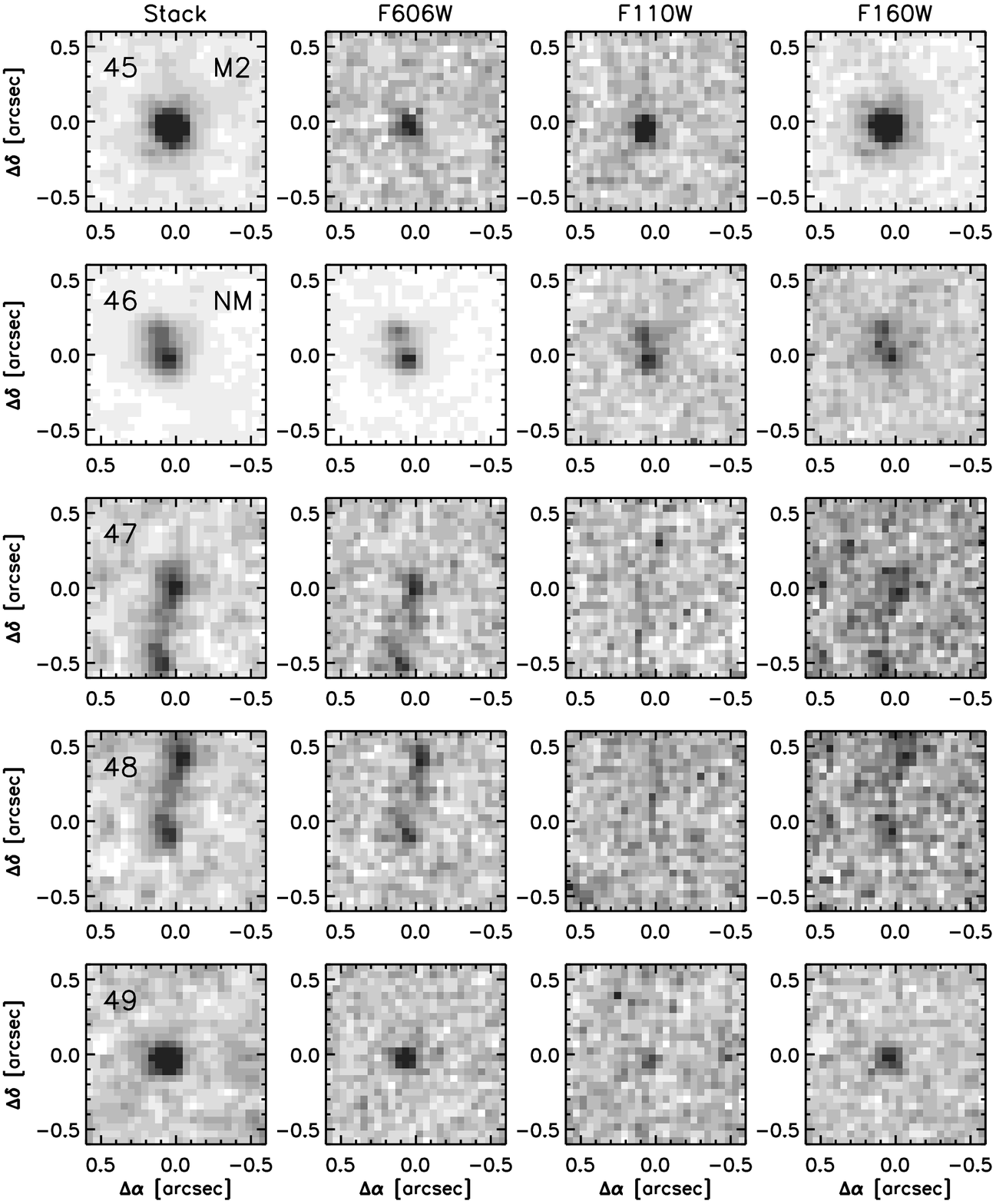}
\caption[Postage Stamps]
{Postage stamp images of the compact sources in the vicinity of LABd05.
}
\label{fig:imappendix5}
\end{figure}

\begin{figure}
\center
\includegraphics[angle=0,width=6.5in]{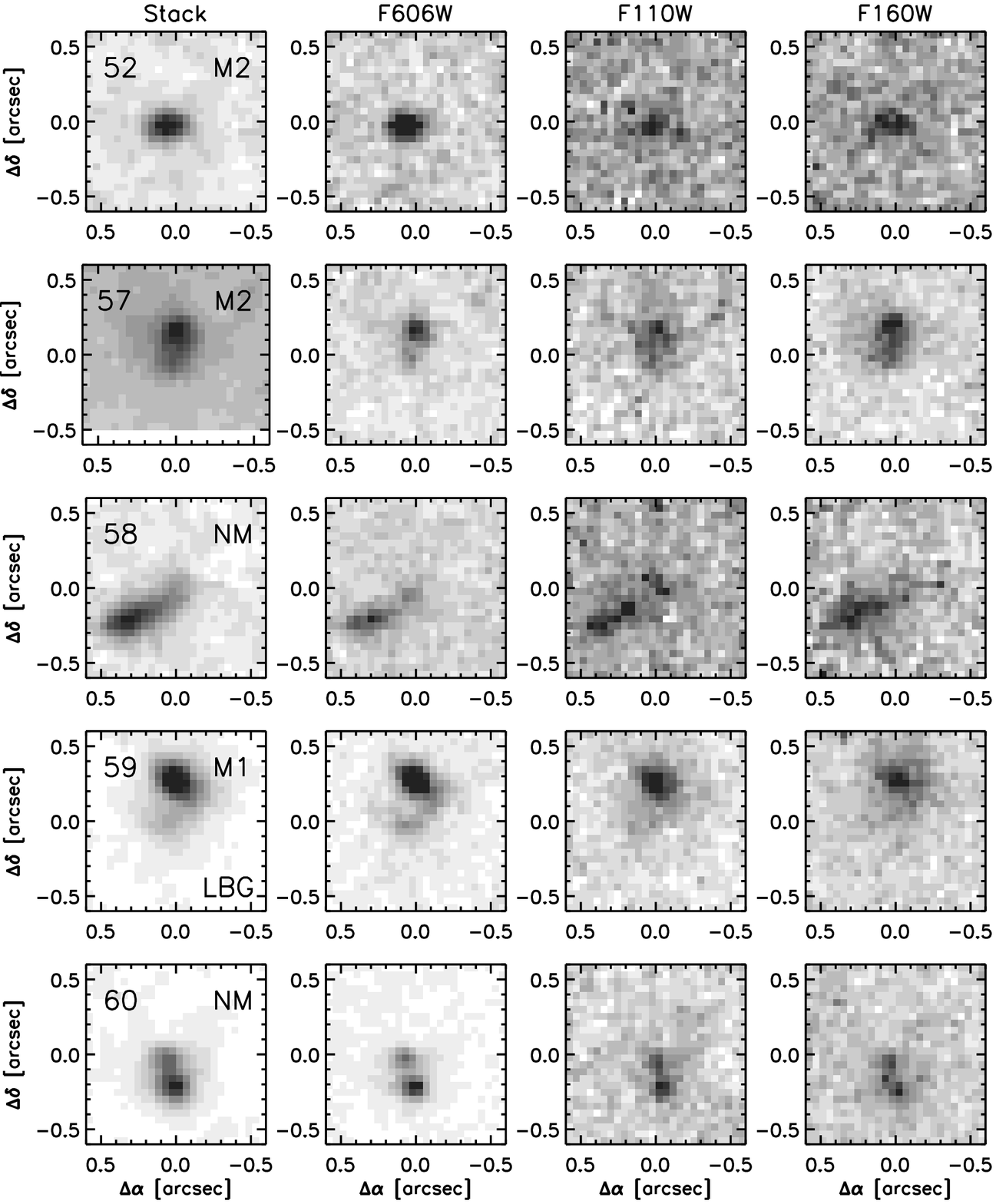}
\caption[Postage Stamps]
{Postage stamp images of the compact sources in the vicinity of LABd05.
}
\label{fig:imappendix6}
\end{figure}

Postage stamp images of the compact sources in the vicinity of LABd05 are 
shown in Figures~\ref{fig:imappendix1}-\ref{fig:imappendix6}.  From left to right, the panels 
display the stacked image and the individual $V_{606}$, $J_{110}$, and $H_{160}$-band images.  
Object ID numbers are shown in the upper left-hand corner of the stacked image, and the membership category 
(where applicable) is given in the upper right-hand corner.  The AGN (\#\agnid) and both components of 
the LBG (\#\lbgidA\ and \#\lbgidB) are labeled in the lower right-hand corner. 

We note that the current data are not sufficient to distinguish whether the object pairs 
(\#\lbgidA+\lbgidB, \#1+57, \#21+58, and \#\interidA+\interidB) 
are true companions or just morphological peculiarities (i.e., tidal features, dust lanes, etc.) 
associated with the primary object.  
For our analysis we have chosen to treat each of the object pairs 
as two separate objects.  Treating them each as a single object would 
decrease the number of sources in the ``M1" category by one and the 
``M2" category by two, but this does not significantly alter our 
conclusions regarding the nature of the member galaxies or the 
overall luminosity function within the LABd05 system.

\clearpage

\end{document}